\documentclass[epj]{svjour}

\usepackage{graphics}
\usepackage{amsmath}
\usepackage{amssymb}
\usepackage{hyperref}
\newcommand{\beq}{\begin{equation}}
\newcommand{\eeq}{\end{equation}}

\newcommand{\Eq}[1]{eq.~(\ref{#1})}
\newcommand{\Fig}[1]{fig.~\ref{#1}}
\newcommand{\Sec}[1]{sect.~\ref{#1}}
\newcommand{\ie}{\textit{i.e. }}
\newcommand{\eg}{\textit{e.g. }}

\begin{document}
\title{
Inhomogeneous chiral symmetry breaking in dense neutron-star matter}
\author{Michael Buballa\inst{1} \and Stefano Carignano\inst{1}\thanks{Present address: INFN, Laboratori Nazionali del Gran Sasso, Assergi (AQ), Italy.}}
\institute{Theoriezentrum, Institut f\"ur Kernphysik, Technische Universit\"at Darmstadt, Germany }
\date{Received: date / Revised version: date}
%
\abstract{
An increasing number of model results suggests that chiral symmetry is broken inhomogeneously 
in a certain window at intermediate densities in the QCD phase diagram.
This could have significant effects on the properties of compact stars, possibly leading to new astrophysical signatures.
In this contribution we discuss this idea by reviewing recent results on inhomogeneous chiral symmetry breaking 
under an astrophysics-oriented perspective. 
After introducing two commonly studied  spatial modulations of the chiral condensate, 
the chiral density wave and the real kink crystal,
we focus on their properties and their effect on the equation of state of quark matter. We also describe how these crystalline phases are affected by different
elements which are required for a realistic description of a compact star, such as charge neutrality, the presence of magnetic fields, vector interactions and the interplay with color superconductivity.  Finally, we discuss possible signatures of inhomogeneous chiral symmetry breaking in  the core of compact stars,
considering the cases of mass-radius relations and neutrino emissivity explicitly.
} 
\maketitle
\section{Introduction: inhomogeneous phases in dense quark matter }
\label{intro}

Compact stars provide the only known realization of ultra-dense matter in nature. 
While the external layers of these stars are expected to be formed by 
more conventional forms of matter, like nuclei, neutrons, protons, and electrons, 
the composition of matter in the center and its properties are still unresolved.
This in turn implies that observations of compact stellar objects might help determine the properties of strong interactions at finite densities, which is one of the most challenging tasks of contemporary nuclear physics. While ab-initio lattice calculations are still unable to provide definite answers, the phase structure  in this region
might be extremely rich. 
Already half a century ago, it was speculated that the central densities of compact stars could be high enough 
to liberate quark degrees of freedom \cite{Ivanenko:1965dg,Itoh:1970uw}.
More recently this idea was challenged by the discovery of two compact objects of two solar 
masses~\cite{Demorest,Antoniadis}, requiring a rather stiff equation of state. 
This puts severe constraints on the appearance of quark matter 
in compact stars, but does not rule out this possibility~\cite{Alford:2006vz} (see 
\cite{Alford:2015gna} in this topical issue for a detailed analysis and
\cite{Buballa:2014jta} for an overview).

In this contribution we assume that quark matter is present in compact stars and discuss its properties
and possible phenomenological consequences.
In particular, in the past decade several studies have suggested that dense quark matter might form 
crystalline, \ie spatially inhomogeneous structures, leading to a significant revision of the current picture 
of the phase diagram of quantum chromodynamics.  
These structures are related to the condensation of fermions
 into pairs carrying a finite momentum, translating into an explicit spatial dependence of such condensates.

Prominent examples of inhomogeneous phases in dense quark matter are related to the phenomena of color super\-conductivity and chiral symmetry breaking.  
Inhomogeneous color superconductors might appear in isospin imbalanced quark matter, where the separation between the quark Fermi surfaces makes the formation of diquark pairs with finite momentum the energetically favored channel. Inhomogeneous chiral condensates on the other hand arise in the region 
between the low-density phase where chiral symmetry is spontaneously broken and the high-density restored one.  In the usual picture limited to homogeneous chiral condensates made of quark-antiquark pairs, 
the presence of a chemical potential induces a stress on the formation of these pairs, eventually leading to chiral restoration. On the other hand, at finite densities the formation of pairs made of quarks and holes at the Fermi surface becomes a competitive condensation channel. Since pairs with large relative momenta are disfavored, the preferred pairing mechanism leads to a condensate with a non-vanishing net total momentum, \ie a spatially inhomogeneous one (for a more detailed discussion, see \eg \cite{Kojo:2009}). 

\begin{figure}
\resizebox{0.47\textwidth}{!}{%
  \includegraphics{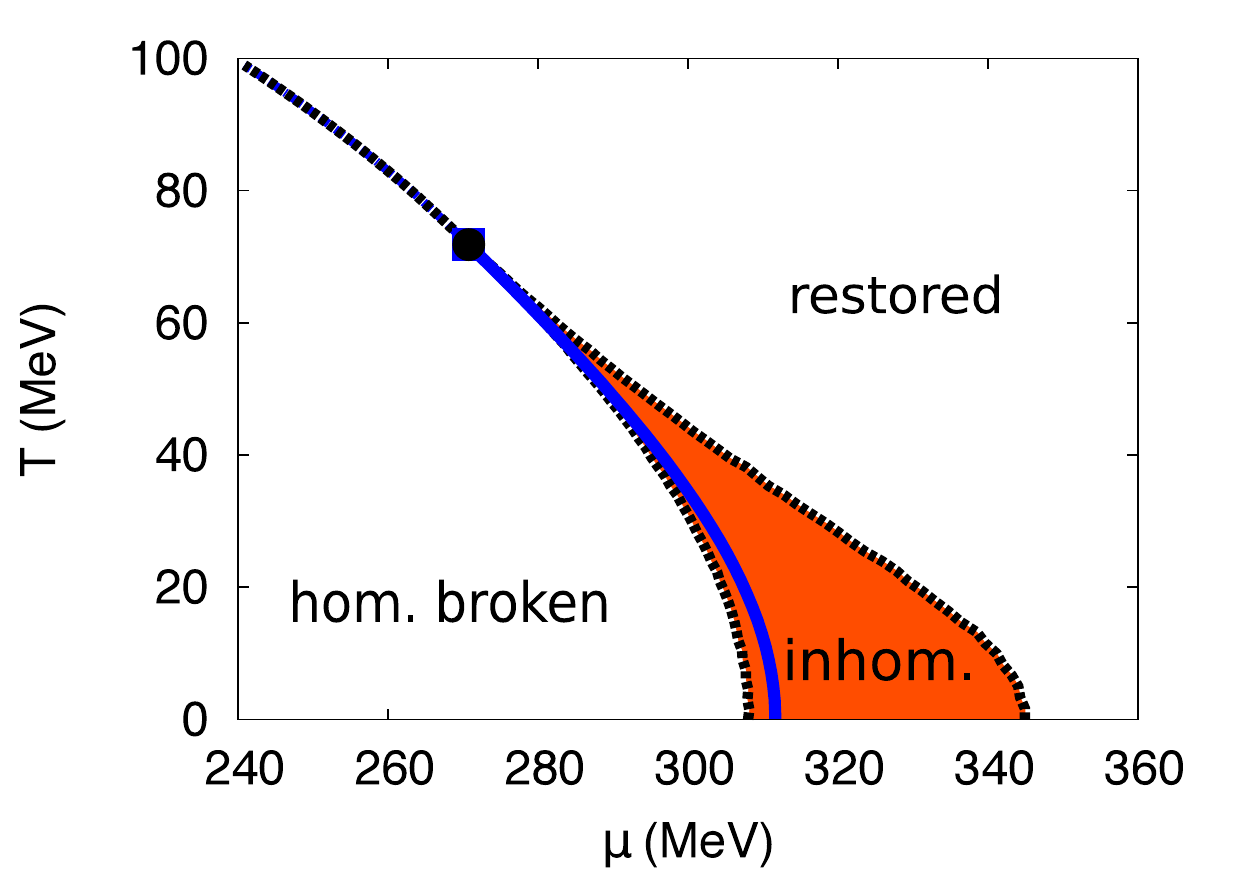}
}
\caption{Phase diagram in the $\mu$-$T$ plane.
The shaded region indicates the inhomogenous phase, covering the first-order phase boundary 
(blue solid line) between the homogeneous chirally broken and restored phases. 
This phase diagram is a reproduction of the result shown in ref.~\cite{Nickel:2009wj}
and was obtained within an NJL model with the RKC ansatz, see \Sec{sec:model}.
}
\label{fig:pd}       
\end{figure}

The theoretical investigation of inhomogeneous condensates is mostly performed through effective models,
most notably the Nambu--Jona-Lasinio (NJL) and  Quark-Meson models.
Most studies of this kind find
 an inhomogeneous ``island'' at low temperatures and finite densities, like in the phase diagram
depicted in \Fig{fig:pd}: crystalline chiral condensates are thermodynamically favored in a finite chemical potential window, 
typically covering the usual first-order phase transition which is expected to occur for homogeneous matter.
More recently, this picture was confirmed by a Dyson-Schwinger calculation, directly based on QCD~\cite{Muller:2013tya}.
The actual extension of 
the inhomogeneous window is still uncertain: 
the Dyson-Schwinger and some NJL studies
suggest that, at least at low temperatures, inhomogeneous chiral symmetry breaking might occur in a very wide range of densities \cite{Muller:2013tya,CB:2011}, while other models predict a smaller size. In this sense, \Fig{fig:pd} could be seen as a rather conservative estimate. 

At any rate, the realization of inhomogeneous chiral symmetry breaking appears to be a rather robust prediction for the region of the phase diagram which is expected to be relevant for the description of compact stars. 
If quark matter is present in the core of compact stars, the formation of crystalline condensates might therefore have significant consequences on the physical properties of these objects, possibly leading to new signatures. 
Since possible observables related to crystalline color superconductivity have already been discussed in \cite{Anglani:2013gfu}, here we will mainly focus on inhomogeneous chiral condensates. 
Also, since
our main objective is to discuss the relevance of inhomogeneous quark matter for compact stars properties,
we will mostly consider vanishing temperatures and try to present established results under a new, more astrophysics-oriented perspective. 
The interested reader can find a complementary review with a more general discussion on inhomogeneous chiral condensates, including their finite-temperature properties and resulting phase structures in \cite{Buballa:2014tba}
(see also \cite{Broniowski:2011} for a brief historical review).

In the following, after describing basic features of inhomogeneous chiral symmetry breaking and its effects on some thermodynamical quantities of a dense quark system, we will discuss how the physical conditions realized inside compact stars, such as charge neutrality or the presence of strong magnetic fields, might affect the properties of these crystalline  condensates.

\section{Properties of inhomogeneous chiral symmetry breaking}

\label{sec:model}

In this section we want to give a brief overview about the basic features of inhomogeneous 
chiral condensates and how they are formally derived in model calculations.
To be specific, we consider an NJL model, defined by the Lagrangian 
\beq
\mathcal{L}_\text{NJL} =
 \bar{\psi}\left(i\gamma^\mu \partial_\mu-m\right)\psi +
G\left(\left(\bar{\psi}\psi\right)^2+\left(\bar{\psi}i\gamma^5\tau^a\psi\right)^2\right),
\label{eq:NJL}
\eeq
where $\psi$ denotes a quark field with bare mass $m$,  
interacting by local four-point vertices proportional to the dimensionful coupling constant $G$.	
For simplicity, we begin by considering here the standard two-flavor model with scalar-isoscalar and pseudoscalar-isovector 
interactions~\cite{NJL2}, while we will discuss various model extensions in later sections.

In order to analyze the thermodynamic  properties at temperature $T$ and quark chemical potential 
$\mu = \mu_B/3$ ($\mu_B$ being the baryon chemical potential)
we need to calculate the grand potential per unit volume $V$,
\beq
\Omega(T,\mu) 
=
-\frac{T}{V} \log\mathcal{Z}(T,\mu),
\eeq
where $\mathcal{Z}$ denotes the grand canonical partition function.
To this end, we perform a mean-field approximation in the presence of the scalar and pseudoscalar condensates 
\beq
 S({\bf x})  = \langle\bar{\psi}\psi\rangle 
 \quad \text{and} \quad   
P({\bf x}) = \langle\bar{\psi}i\gamma^5\tau^3\psi\rangle,
\label{eq:SP}
\eeq
respectively. 
Here we assume that only the third isospin component of the pseudoscalar condensate is non-vanish\-ing.
 By this,
we exclude the possibility of charged pion condensation
which at any rate is not expected to occur in compact stars, where the isospin chemical potential
stays well below the pion mass (\Sec{sec:neut}).

We also assume that the condensates are time independent.
However, since we are interested in inhomogeneous phases, we must retain their dependence 
on the spatial coordinate ${\bf x}$. 
As a consequence, the derivation of the mean-field thermodynamic potential
becomes significantly more difficult than for space-independent condensates.
As discussed in detail in ref.~\cite{Buballa:2014tba}, one finds
\beq
\Omega_\text{MF}(T,\mu; S, P) 
=
\Omega_\mathit{kin} + \Omega_\mathit{cond}
\,,
\eeq
with the condensate part
\beq
\Omega_\mathit{cond}
= 
\frac{1}{V}\int_{V} d^3x\,  G\left(S^2({\bf x}) + P^{\,2}({\bf x})\right) 
\eeq
and the kinetic part
\beq
\Omega_\mathit{kin} 
=
 -\frac{1}{V} \sum\limits_\lambda \left[ \frac{E_\lambda -\mu}{2} + T \log\left(1 + e^{-\frac{E_\lambda -\mu}{T}}\right)\right]\,,
\eeq
where $\lambda$ labels the eigenstates of the effective Dirac Hamiltonian
\begin{equation}
{H}({\bf x})  =  \gamma^0 \left[ -i\gamma^i\partial_i + m - 2G(S({\bf x}) + i \gamma^5 \tau^3  P({\bf x})) \right],
\end{equation}
and $E_\lambda$ are the corresponding eigenvalues.
In contrast to the homogeneous case, the diagonalization of $H$ for arbitrary space-dependent condensate functions 
is rather difficult and must in general be performed numerically. 
After that, $\Omega_\text{MF}$ must be minimized with respect to the functions $S({\bf x})$ and $P({\bf x})$ in order to find the 
energetically favored ground state. 
Since this problem has not yet been solved in $3+1$ dimensions, 
one considers simple ans{\"a}tze for the condensate functions.
This reduces the problem to finding the minimum of $\Omega_\text{MF}$
with respect to a limited set of variational parameters. 
Most parametrizations used in literature correspond to one-dimensional spatial modulations. 
Two- and higher-dimensional modulations have been studied as well, 
but (unlike for color superconductors) were found to be disfavored against one-dimensional ones, at least within the mean-field approximation~\cite{Abuki:2011,Carignano:2012sx}.

In the following we will restrict ourselves to the chiral limit ($m=0$) and
consider two one-dimen\-sional ans\"atze which even allow for an analytical diagonalization of $H$
and are therefore the most commonly used in the literature.
Choosing the modulation to be along the $z$-direction and introducing the complex mass function
\beq
M(z) = -2G \big(S(z) + i P(z) \big) ,
\label{eq:Mx}
\eeq
the most simple (and therefore most popular) one is the so-called (dual) chiral density wave (CDW)
\cite{Broniowski:1990,NT:2004}
\beq
M(z) = \Delta e^{i qz} ,
\label{eq:CDW}
\eeq
\ie a single plane wave with amplitude $\Delta$ and wave number $q$, corresponding to a spatial period
of $L = 2\pi/q$.
When moving along the $z$-direction,  the CDW rotates around the chiral circle, 
transforming scalar condensates into pseudoscalar ones and vice versa:
\beq
  S(z)=-\frac{\Delta}{2G} \cos{(qz)} \,,\quad P(z)=-\frac{\Delta}{2G} \sin{(qz)} . 
\label{eq:CDWcond}
\eeq

A more sophisticated ansatz is the ``real kink crystal'' (RKC), corresponding to an array of domain-wall 
solitons.
It was derived from the analytically known solutions of the $1+1$ dimensional 
Gross-Neveu model~\cite{Schnetz:2004}, embedding them into $3+1$ dimensions 
by making use of Lorentz boosts into the transverse directions \cite{Nickel:2009wj}.
Although there is no general proof,
 the RKC modulations have so far turned out to be the energetically most
favored ones among all ans\"atze considered in the literature, including higher-dimensional ones \cite{Abuki:2011,Carignano:2012sx},
so that to our current knowledge they consitute the most likely shape realized in the 
inhomogeneous island. 

The mass function reads
\beq
M(z) = \Delta \sqrt{\nu} \, \mathrm{sn}(\Delta z | \nu)       
\eeq
and depends on two parameters, $\Delta$ and $\nu$.
Here $\mathrm{sn}(\alpha|\nu)$ denotes a Jacobi elliptic function with elliptic modulus $\nu \in [0,1]$, interpolating between 
$\tanh(\alpha)$ for $\nu = 1$ and $\sin(\alpha)$ for $\nu = 0$. 
The amplitude of the modulation is then given by $\Delta \sqrt{\nu}$
and its spatial period by $L = 4\mathbf{K}(\nu)/\Delta$, where $\mathbf{K}(\nu)$ is a complete elliptic integral of the first kind.
In contrast to the CDW, the RKC is real (hence the name), \ie the pseudoscalar condensate vanishes identically.

\begin{figure*}
  \resizebox{0.32\textwidth}{!}{%
  \includegraphics{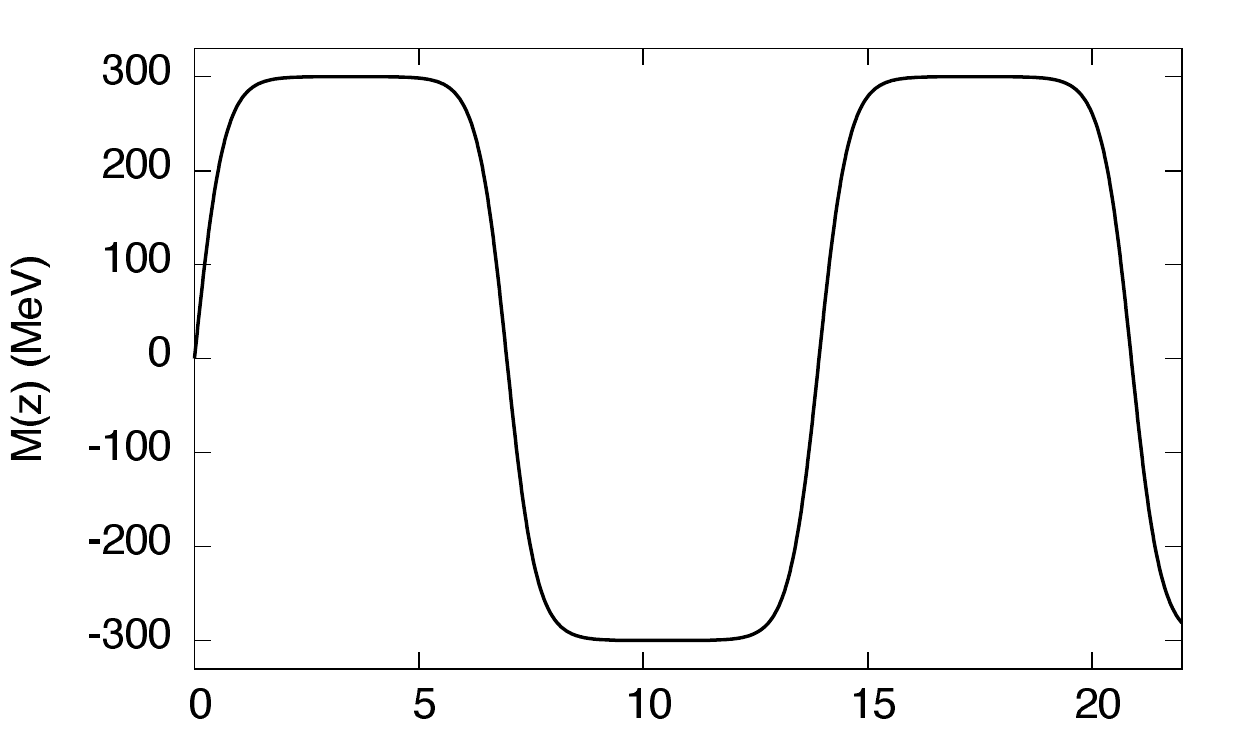} 
  }
  \resizebox{0.32\textwidth}{!}{%
  \includegraphics{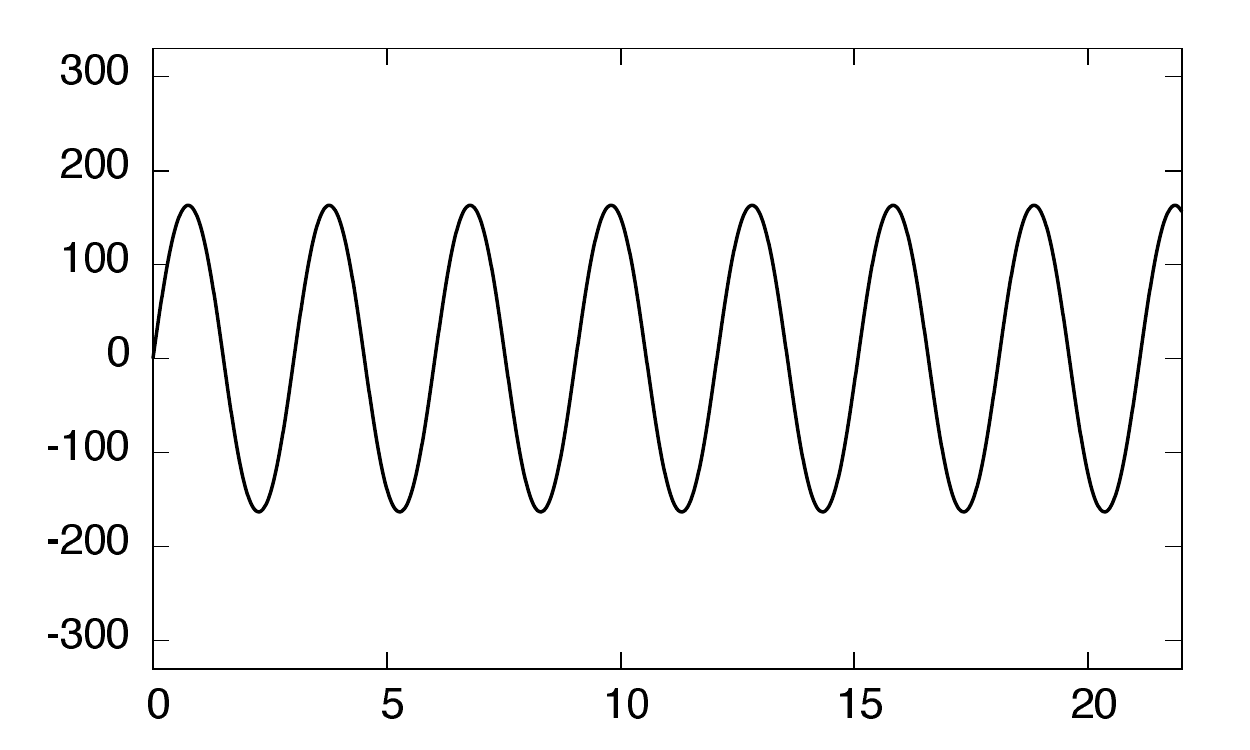} 
  }
   \resizebox{0.32\textwidth}{!}{%
  \includegraphics{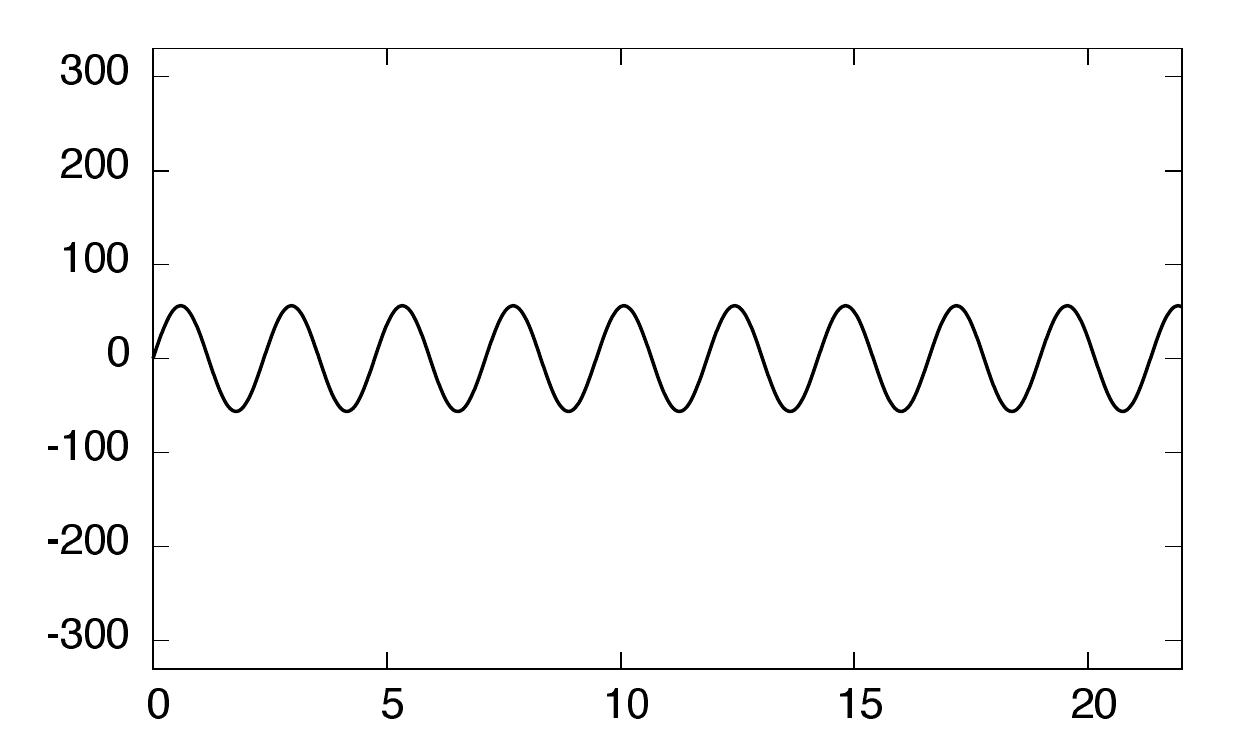} 
  }
  
\resizebox{0.328\textwidth}{!}{%
  \includegraphics{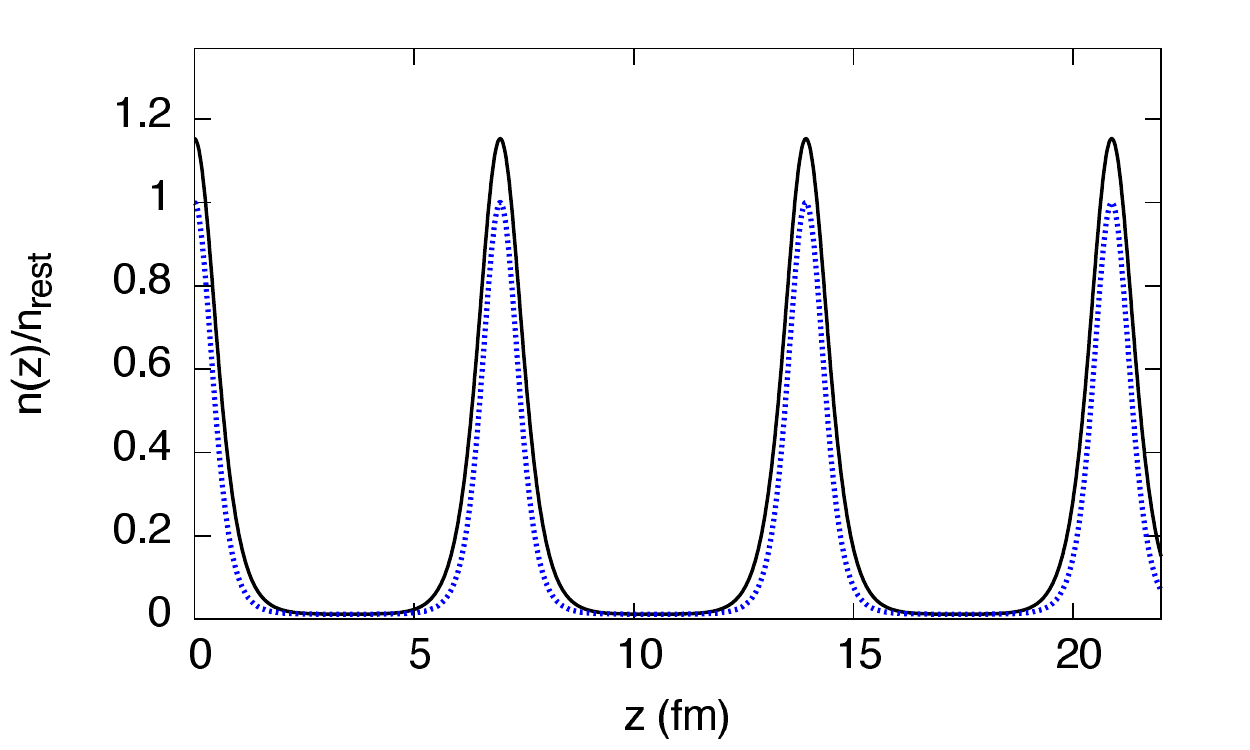} 
  }
  \hspace{-.25cm}
  \resizebox{0.328\textwidth}{!}{%
  \includegraphics{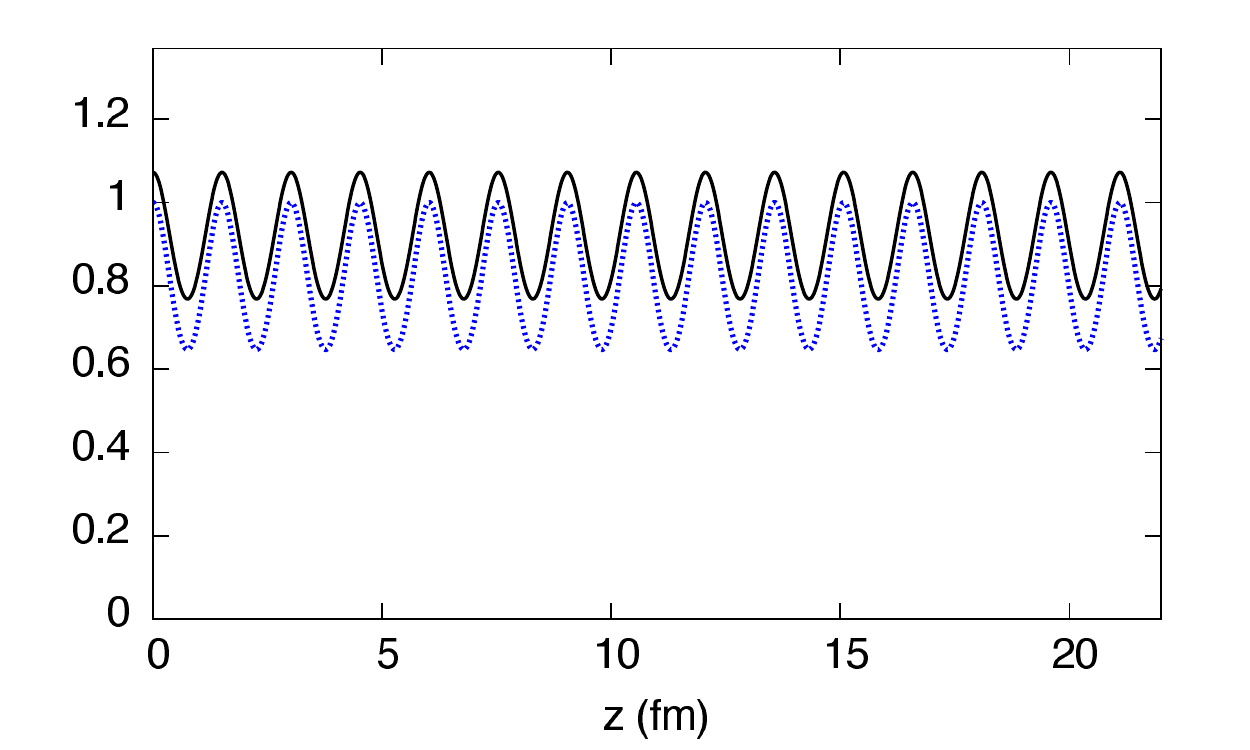} 
  }
  \hspace{-.25cm}
   \resizebox{0.328\textwidth}{!}{%
  \includegraphics{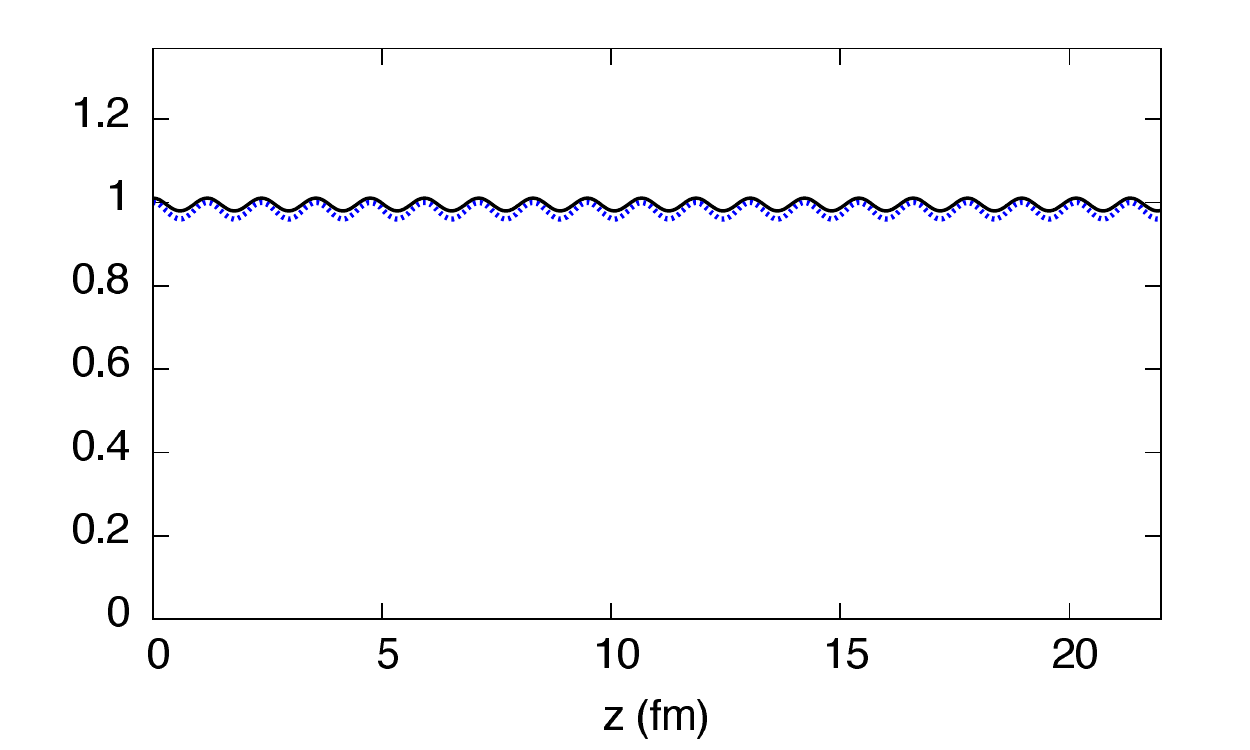} 
  }
  
\caption{Mass functions (upper row) and density profiles normalized to the density of chirally restored matter (lower row) for an RKC modulation, evaluated at zero temperature 
and $\mu = 308$ (left), 325 (center) and 345 MeV (right), 
corresponding to regions  near the phase boundary to the homogeneous chirally broken phase, 
in the center of the inhomogeneous phase and close to the phase boundary to the restored phase,
respectively. 
For the density, the actual RKC results (solid lines) are compared with the ones obtained in the LFG approximation (dotted lines).
}
\label{fig:RKCshapes}   
\end{figure*}

\begin{figure}
\resizebox{0.47\textwidth}{!}{%
  \includegraphics{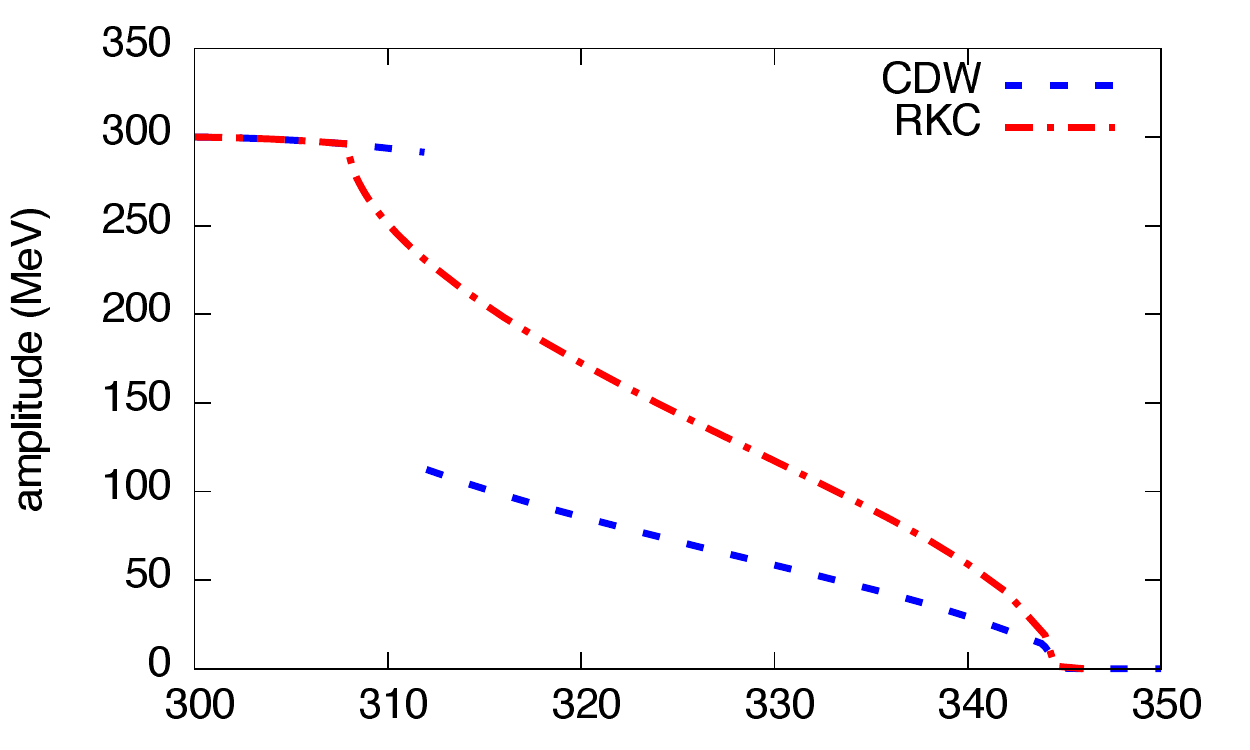} 
   }
 \resizebox{0.47\textwidth}{!}{%
   \hspace{4mm}
 \includegraphics{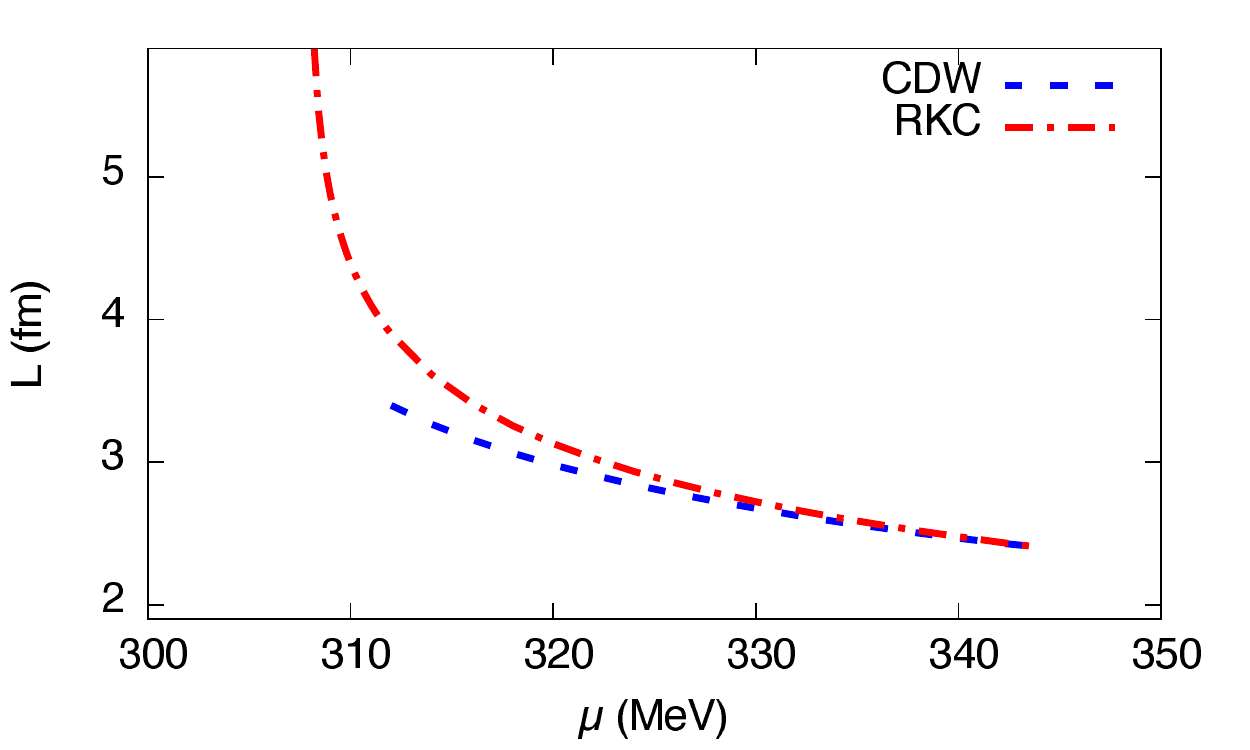} 
 }
  \caption{Amplitudes  (upper panel) and spatial periods (lower panel) of the mass functions $M(z)$, allowing for CDW (dashed)
  or RKC (dash-dotted) modulations, as functions of the quark chemical potential at $T=0$.}
\label{fig:ampper}  
\end{figure}

Examples for the shape of this mass modulation are displayed in \Fig{fig:RKCshapes}.
In the lower part of the figure we also show the corresponding density profiles, which have been derived in ref.~\cite{CNB:2010}.
As one can see, the structure is most pronounced 
near the phase boundary to the homogeneous phase with broken chiral symmetry.  
Here the density is sharply peaked around the 
zero crossings of the mass function, while it is suppressed in the regions where $|M(z)|$ is large. 
Increasing the chemical potential,
the mass amplitude decreases (see also \Fig{fig:ampper}) and the density profiles become more and more washed out,
smoothly approaching the corresponding density  in the restored phase.

This observed behavior suggests us that the shape of the density is largely determined by the mass function.
In order to quantify this, 
we introduce a
 local Fermi gas (LFG) approximation, where the density at each point in space is calculated
by the corresponding expression for a homogeneous ideal Fermi gas of quarks with mass $M(z)$,
\ie  at $T=0$,
\beq
\label{eq:LFG}
        n_\mathit{LFG}(z) = \frac{N_f N_c}{3\pi^2} \left(\mu^2 - |M(z)|^2 \right)^{3/2} \, \theta(\mu - |M(z)|).
\eeq
The resulting densities are shown in the lower panel of \Fig{fig:RKCshapes} as dotted lines.
Comparing them with the exact mean-field results from \cite{CNB:2010} (solid lines)
we can see that, although the two share a qualitatively similar behavior, the actual density for the RKC solutions is 
always larger than the LFG one. 
In particular, the RKC density overshoots the density for the restored phase \cite{CNB:2010,Buballa:2012vm}, while 
$n_\mathit{LFG}/n_\mathit{rest}$ is always $\leq 1$, as can easily be seen from \Eq{eq:LFG}.

While for the RKC the density is spatially varying, following the changes in the mass function, 
 in the case of a CDW modulation, for which $|M(z)| = \Delta = \mathit{const.}$, the density is constant in space.
Nevertheless, similar to the RKC case, the density for the CDW  is larger than the one obtained for a 
spatially homogeneous order parameter $M = \Delta$.
 
In the numerical examples presented in this section, we employed a Pauli-Villars type regularization (for more details, see \cite{Nickel:2009wj}), and
 the model parameters are fitted to 
reproduce the chiral-limit value $f_\pi = 88$~MeV of the pion decay constant 
and a constant constituent quark mass of 300~MeV in vacuum. Another popular regularization scheme, which is often 
used
when dealing with magnetic fields, is the Schwinger proper time method, for which the parameters are usually fitted to a slightly larger constituent mass of 330~MeV, see \eg \cite{NT:2004,Frolov:2010}. We note in any case that the choice of a different regularization does not qualitatively alter the results (with the notable exception of a sharp three-momentum cutoff, which is not appropriate when dealing with inhomogeneous condensates since quark momenta are not conserved quantities).

Minimizing the thermodynamic potential for the RKC ansatz
with respect to $\Delta$ and $\nu$ at given values of $\mu$ and $T$ then
gives rise to the phase diagram shown in \Fig{fig:pd}, which we already discussed in the Introduction.
Using the CDW ansatz and minimizing with respect to $\Delta$ and $q$ yields a rather similar phase diagram.
Although the CDW is disfavored against the RKC \cite{Nickel:2009wj},
working with the simpler CDW ansatz is therefore often sufficient to get a rough picture of the phase structure for inhomogeneous phases.

The main qualitative difference between the two modulations concerns the transition from the homogeneous broken to the inhomogeneous phase,
which is first order for the CDW and second order for the RKC.
This can also be seen in \Fig{fig:ampper}, where the amplitudes and periods are displayed for both modulations.  
At the phase transition, the amplitude drops discontinuously for the CDW ansatz, while it is continuous for the RKC.
For the latter the period $L$ diverges at this point, enabling a smooth transition to the homogeneous phase.
The phase transition from the inhomogeneous to the restored phase, on the other hand, 
is second order in both cases, with the amplitudes smoothly going to zero.
  
Except for the region close to the phase transition to the homogeneous broken phase,
the periods are almost identical for both ans\"atze, roughly being of the order  of 2~--~3~fm.
The amplitudes, on the other hand, differ by about a factor of 2.
These results are consistent with predictions of a Ginzburg-Landau analysis near the second-order transition 
to the restored phase~\cite{Nickel:2009ke,Abuki:2011,Buballa:2014tba}.

\begin{figure}
\resizebox{0.47\textwidth}{!}{%
  \includegraphics{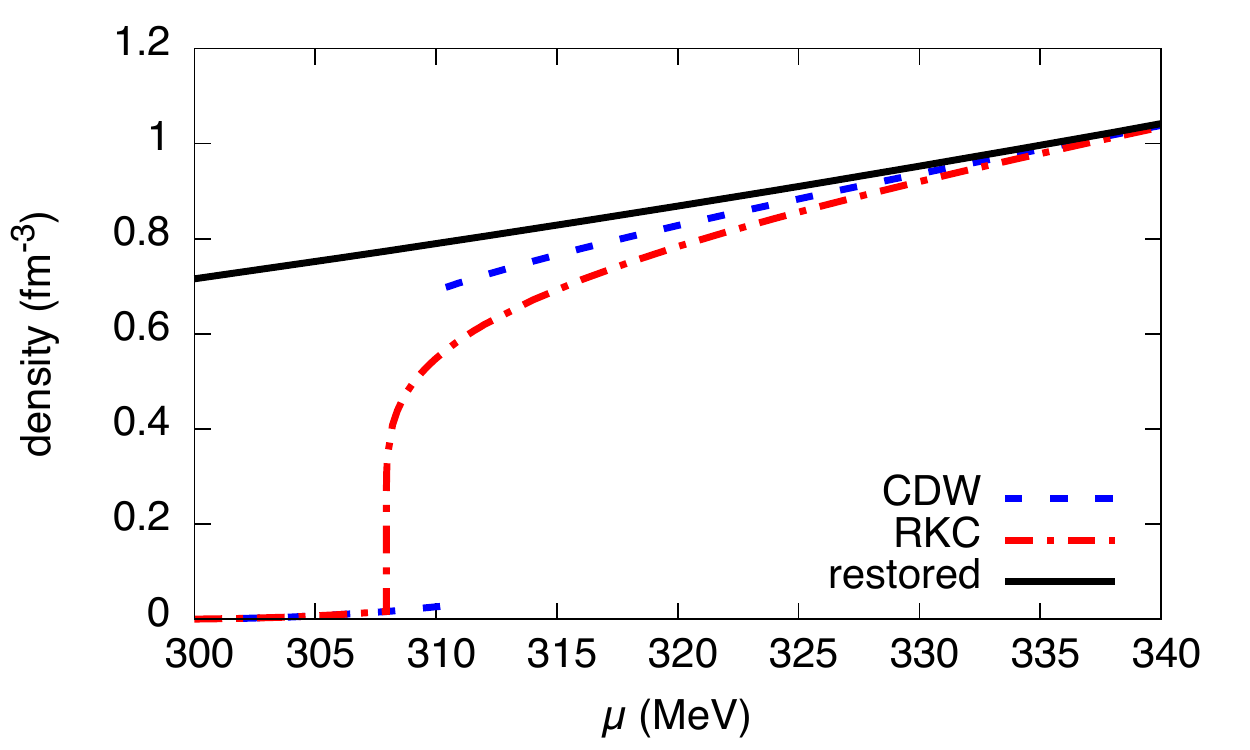} 
  }
  \caption{Space-averaged quark-number density $\bar{n}$ as a function of the quark chemical potential at $T=0$.
The dashed and dash-dotted lines indicate the results obtained when allowing for CDW and RKC modulations,
respectively. The density for the restored solution ($M=0$) is indicated by the solid line.
Note that the baryon number densities $\bar{n}_B$ are related to the quark ones by $\bar{n}_B = \bar{n}/3$.
 }
\label{fig:density}  
\end{figure}

The spatially averaged quark-number densities 
\beq
        \bar{n} = -\frac{\partial\Omega}{\partial\mu}
\eeq
are displayed in \Fig{fig:density}, together with the corresponding density for the restored solution.  
At fixed chemical potential, $\bar{n}$ is smallest for the RKC and largest for the restored solution, with the CDW lying
in between. 
Again, this can qualitatively be explained as a suppression effect due to the effective quark mass, 
which vanishes in the restored phase and is largest in the RKC.\footnote{This remains true when we take the spatial
average of $|M(z)|^2$.
}

For the RKC ansatz, the inhomogeneous phase starts essentially at $\bar{n} = 0$, corresponding to infinitely separated
domain-wall solitons.
For the CDW solution, on the other hand, the minimum density is around $1.4~n_0$, 
where $n_0 \approx 0.5$~fm$^{-3}$ is the quark number density in symmetric nuclear matter at saturation.
In both cases we find the  upper density of the inhomogeneous island to be about $2.2~n_0$.
Keeping in mind that these numbers are model and parameter dependent and could also be higher,
this lies in the ballpark of densities expected in a possible quark phase in compact stars.

\begin{figure}
\resizebox{0.47\textwidth}{!}{%
  \includegraphics{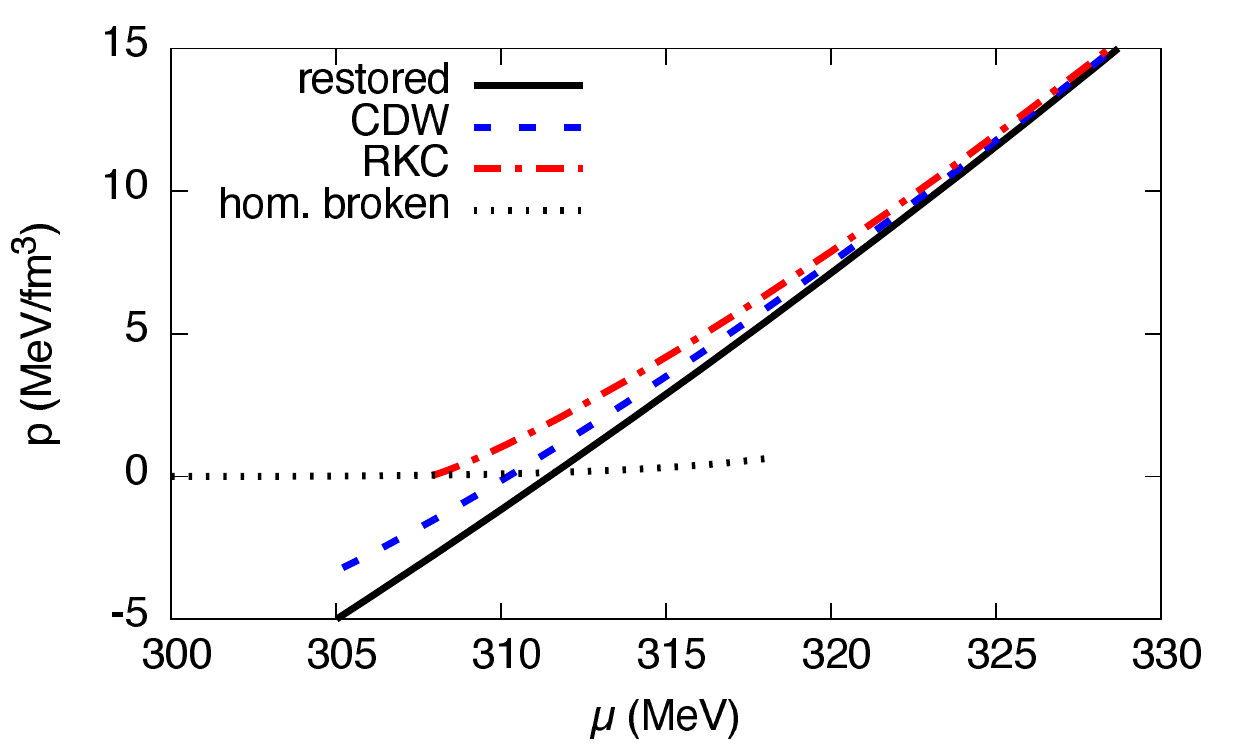} 
  }
  \caption{Pressure at $T=0$ as a function of the quark chemical potential for different mass functions.
  Solid: restored solution, dotted: homogeneous broken solution, dashed: CDW, dash-dotted: RKC.   
  }
\label{fig:pressure}  
\end{figure}

The pressure $p = - (\Omega- \Omega_\mathit{vacuum})$ 
 of the various solutions at $T=0$ as function of $\mu$ is shown 
in \Fig{fig:pressure}.  As usual, the favored solution is the one with the largest pressure. 
At low chemical potentials, this is the homogeneous broken solution (dotted line).
Excluding the possibility of inhomogeneous phases, there would be a first-order phase transition to the 
restored phase (solid line), characterized by the crossing of the two lines at the critical chemical potential.
However, by allowing for inhomogeneous solutions, we see that in this regime the RKC (dash-dotted line),
which sets on in a second-order phase transition at a slightly lower chemical potential, is more favored.
Additionally, we observe that the CDW is always disfavored against the RKC, but still favored over the homogeneous solutions in most of this region.
At higher chemical potentials the pressure curves of the inhomogeneous solutions approach the one
of the restored phase and eventually reach it in a second-order transition (not shown in the figure). 

\begin{figure}
\resizebox{0.47\textwidth}{!}{%
  \includegraphics{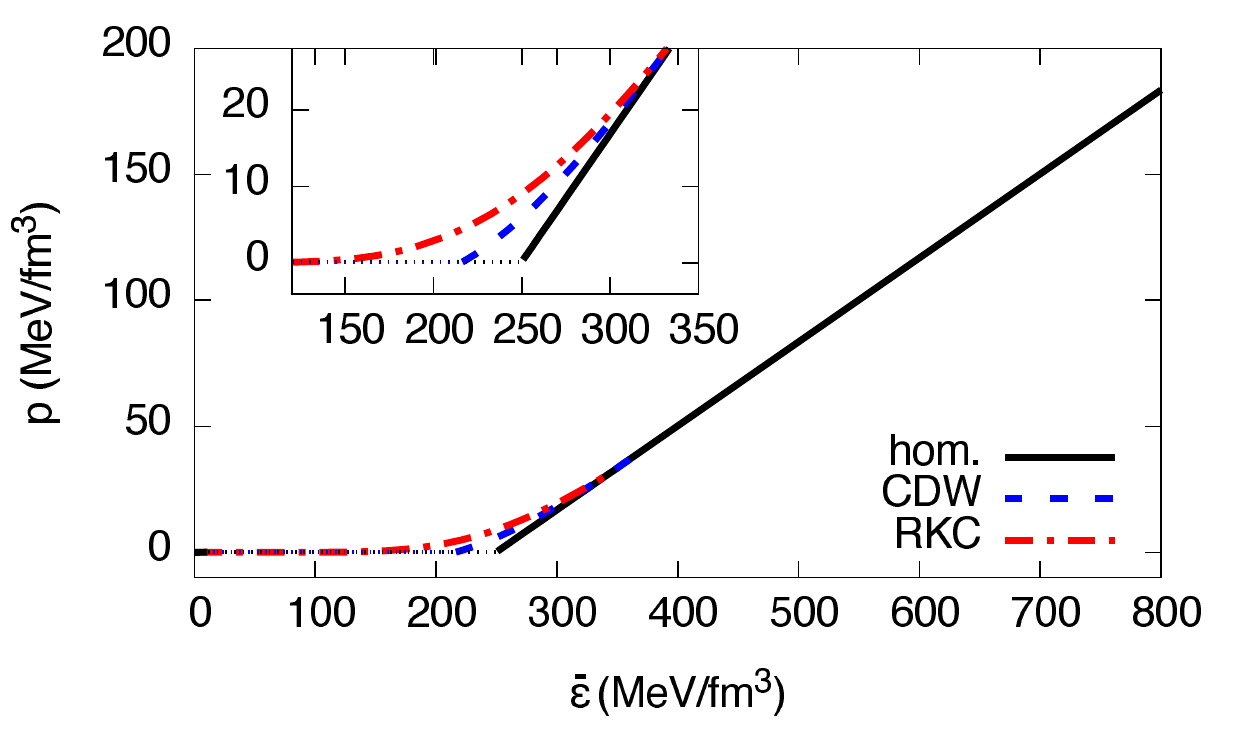} 
  }
  \caption{Equation of state (pressure $p$ vs.\@ spatially averaged energy density $\bar{\varepsilon}$) at $T=0$
  for homogeneous matter (solid) and in the inhomogeneous phase with CDW (dashed) or RKC (dash-dotted) modulations.
  All equations of state begin at $p=\bar{\varepsilon}=0$ in the homogeneously broken phase. 
  The dotted horizontal lines indicate discontinuities related to first-order transitions to the restored phase 
  or to the CDW solution.  
  The inset shows an enlarged detail of the transition region.
  }
\label{fig:eos}  
\end{figure}

Finally, in \Fig{fig:eos}, we compare the corresponding equations of state at $T=0$,
\ie the pressure as a function of the spatially averaged energy density, 
\beq
        \bar\varepsilon = -p + \mu \bar{n} .
\eeq
At the first-order phase transitions from the homogeneous broken to the restored phase or the CDW,
$\bar{n}$ (and hence $\bar\varepsilon$) are discontinuous, leading to horizontal jumps in the equation of state. 
The transition to the RKC, on the other hand, is continuous.
However, since the density grows very quickly above the transition point (cf.~\Fig{fig:density}), the resulting 
equation of state is very similar to the discontinuous ones. 
In general, the inhomogeneous equations of state and in particular the RKC one are stiffer than the homogeneous one,
but the differences are rather small. 
We therefore anticipate that the effect of inhomogeneous phases on the mass-radius relation of compact stars
will be negligible. 

\section{Including further elements of compact-star physics}
\label{sec:csp}

After discussing the basic properties of inhomogeneous chiral condensates within a rather simple model,
we now study the effects on them of additional physical conditions encountered in compact stars.

\subsection{Electric neutrality}
\label{sec:neut}

In the previous section we have assumed that up and down quarks share the same chemical potential $\mu$.
 In compact stars, however, one has to  take into account weak decays and ensure global electric neutrality. 
In order to describe this situation, our model must be extended to contain leptons.
Since those are only weakly coupled to the quarks, the total thermodynamic potential 
can be decomposed into a quark and a leptonic part,
\beq
       \Omega_\mathit{tot} = \Omega_q + \Omega_\ell.
\eeq
Moreover, neutrinos can freely leave the star, so that we only need to take into account charged leptons. 

This system is characterized by two conserved global charges: the net quark number and the total electric charge. 
Therefore $\Omega_\mathit{tot}$ depends on two independent 
chemical potentials, the quark number chemical potential $\mu$ and the electric charge chemical potential $\mu_Q$.
The condition for (spatially averaged) electric neutrality is then given by
\beq
       \bar{n}_Q = -\frac{\partial\Omega_\mathit{tot}}{\partial\mu_Q} \overset{!}{=} 0 .
\label{eq:neutral}
\eeq

The chemical potentials of the individual particle speci\-es are given as linear combinations of $\mu$ and $\mu_Q$,
according to their quantum numbers, \ie
\beq
       \mu_u = \mu + \frac{2}{3}\mu_Q, \quad\;
       \mu_d = \mu_s = \mu - \frac{1}{3}\mu_Q      
\label{eq:muudQ}
\eeq
for the quarks and
\beq
      \mu_\ell = -\mu_Q 
\eeq
for the (negatively charged) leptons.

Neglecting Coulomb effects aside from the neutrality constraint, we can approximate the leptonic part of the thermodynamic 
potential by an ideal gas.
The remaining problem is then to calculate the quark part $\Omega_q$ at nonzero $\mu_Q$
and to determine $\mu_Q$ from \Eq{eq:neutral}.

In the quark sector, if we restrict ourselves again to up and down quarks, 
$\mu_Q$ is equivalent to an isospin chemical potential $\mu_I \equiv \mu_u - \mu_d = \mu_Q$.
Inhomogeneous phases in the presence of a nonvanishing isospin chemical potential have been
discussed in refs.~\cite{Abuki:2013vwa,Abuki:2013pla}
within a Ginzburg-Landau analysis, which is expected to be valid near the chiral critical point 
but not reliable at the low temperatures of compact stars.
Recently,  inhomogeneous phases in isospin asymmetric matter have also been studied within the NJL 
model~\cite{Nowakowski:2015ksa}.
Starting from the Lagrangian of \Eq{eq:NJL} a mean-field approximation was performed, 
considering the flavor-diagonal scalar and pseudoscalar condensates 
\beq
 S_f({\bf x})  = \langle\bar{f} f\rangle 
 \quad \text{and} \quad   
P_f({\bf x}) = \langle\bar{f}\,i\gamma^5 f\rangle, 
\label{eq:SfPf}
\eeq
$f \in \{u,d\}$,
and a generalized CDW ansatz of the form 
\beq
  S_f(z)=-\frac{\Delta_f}{4G} \cos{(q_f z)} \,,\quad P_f(z)=-\frac{\Delta_f}{4G} \sin{(q_f z)} .
\label{eq:CDWgen}
\eeq  
Note that the condensates defined in \Eq{eq:SP} are related to the above ones 
by $S=S_u+S_d$ and $P = P_u - P_d$, so that 
\Eq{eq:CDWcond}  is recovered if we choose $\Delta_u = \Delta_d \equiv \Delta$ and $q_u = -q_d \equiv q$.

For $q_u = -q_d$ and arbitrary amplitudes $\Delta_u$ and $\Delta_d$ the effective Dirac Hamiltonian of this ansatz
can still be diagonalized analytically. 
Minimizing the thermodynamic potential for this case it was found that 
the inhomogeneous phase shrinks with increasing $|\mu_I|$, but it still survives at values of $\mu_I$ 
needed to obtain electrically neutral matter~\cite{Nowakowski:2015ksa}.\footnote{Solving \Eq{eq:neutral} 
for this ansatz, $\mu_I$ was found to be  around $70 - 80$~MeV in the inhomogeneous phase at $T=0$.
This is roughly the same result one gets for the restored solution in this regime.
}

It is plausible, however, that the assumption $q_u = -q_d$ is too restrictive:
As we have seen earlier, the spatial periodicity and, hence, the wave number is a function of the chemical potential.
At nonzero $\mu_I$, where the chemical potentials of up and down quarks differ, 
we should therefore expect that unequal values of $|q_u|$ and $|q_d|$ could be favored. 
Unfortunately, for this case the eigenvalues of the Hamiltonian are not known analytically, so that $H$ has to be
diagonalized numerically. 
Obviously, this complicates the calculations considerably. 
In particular, for the numerical diagonalization a fixed ratio $R = q_u / q_d$ of the wave numbers 
must be considered in order to be able to map the momenta onto a discrete grid. 
While in principle any real number can be approximated by a given $R$ and therefore all possible values 
for $q_u$ and $q_d$ can be explored, in practice of course only a few selected ratios can be considered.

\begin{figure}
\resizebox{0.47\textwidth}{!}{%
  \includegraphics{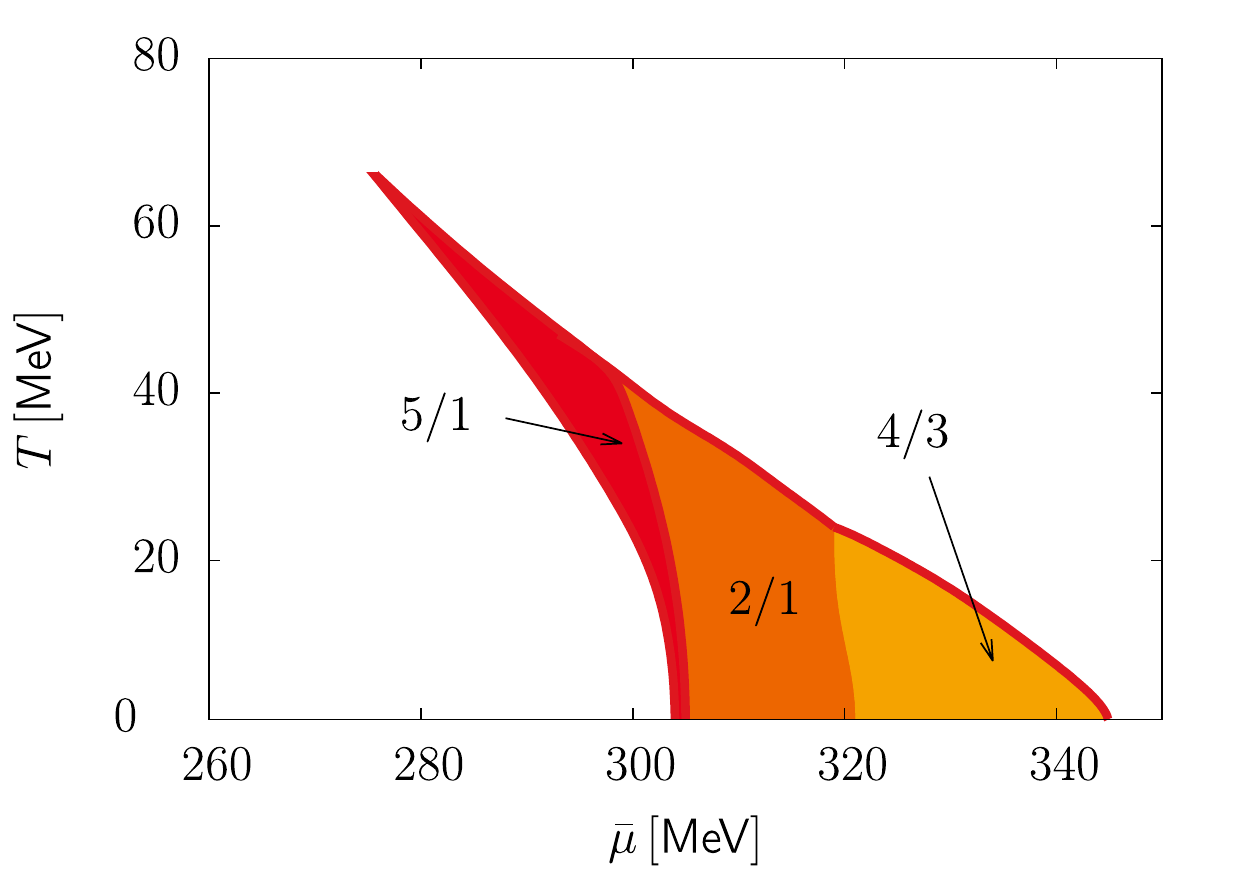}
}
\caption{Phase diagram in the plane spanned by the averaged chemical potential
$\bar\mu = \frac{1}{2}(\mu_u + \mu_d)$ and temperature $T$ at fixed isospin chemical potential
$|\mu_I| = 60$~MeV for the generalized CDW ansatz \Eq{eq:CDWgen}
with $R  = q_u/q_d \in \{-1, \frac{4}{3},2,5\}$.
The shaded area indicates the region where an inhomogeneous solution is favored over homogeneous ones. 
Different colors correspond to different values of the most favored $R$, as indicated by the labels.
The figure was taken from ref.~\cite{Nowakowski:2015ksa}.
}
\label{fig:pdiso}       
\end{figure}

In ref.~\cite{Nowakowski:2015ksa} the analysis was therefore restricted to 
the values $R \in \{-1, \frac{4}{3},2,5\}$, and it was determined which of these ratios is the most favored one
at fixed $|\mu_I| = 60$~MeV.
The resulting phase diagram is shown in 
\Fig{fig:pdiso}.
Here the phase structure is shown in the plane spanned by $\bar\mu = \frac{1}{2}(\mu_u + \mu_d) \equiv \mu + \frac{1}{6} \mu_Q$ 
and the temperature $T$. 
One can see that different ratios are favored in different areas of the inhomogeneous phase, with the general tendency that
$R$ decreases with increasing $\bar\mu$. 
In a more realistic scenario one would of course expect that $R$ varies continuously with $\bar\mu$ and $T$.
However, even within this very limited analysis, we find that the size of the inhomogeneous region at nonvanishing
$\mu_I$ is roughly the same as for $\mu_I = 0$ (cf.~\Fig{fig:pd}), so that this region might become even 
bigger if more ratios are considered. 

So far we restricted ourselves to the standard NJL-model Lagrangian, defined in \Eq{eq:NJL}.
However, for asymmetric quark matter, the exact isospin structure of the interaction becomes 
relevant. 
In particular, \Eq{eq:NJL} can be viewed as a special case of the more general Lagrangian~\cite{Asakawa:1989,Frank:2003ve}
\begin{align}
\mathcal{L}
=
&\phantom{+}\bar{\psi} \big( i\gamma^\mu \partial_\mu - \hat{m} \big) \psi
\nonumber\\
&+ G_1 \Big(    (\bar{\psi}\psi)^2 + (\bar{\psi}\tau^a\psi)^2
                     + (\bar{\psi}i\gamma_5\psi)^2 + (\bar{\psi}i\gamma_5\tau^a\psi)^2  \Big)
\nonumber\\
&+ G_2 \Big(   (\bar{\psi}\psi)^2 - (\bar{\psi}\tau^a\psi)^2
                      - (\bar{\psi}i\gamma_5\psi)^2 + (\bar{\psi}i\gamma_5\tau^a\psi)^2 \Big)
\,,\label{eq:LagNJLiso}
\end{align}
with the current quark matrix $\hat m = \mathrm{diag}_f(m_u,m_d)$ 
and two $SU(2)_L  \times SU(2)_R \times U_V(1)$ symmetric interaction terms, 
proportional to the coupling constants $G_1$ and $G_2$, respectively.
The first term respects an additional $U_A(1)$ symmetry, which is explicitly broken by the second.

Considering again the condensates defined in \Eq{eq:SfPf}, one finds that the 
mean-field Hamiltonian depends on the effective mass operators
\beq
       \hat M_f = m_f - 4G_1 \left(S_f + P_f \,i\gamma^5\right) - 4G_2 \left(S_h - P_h \,i\gamma^5\right) 
\eeq
with the flavor indices $f,h \in \{u,d\}$, $f\neq h$.
From this expression we can see that 
the $G_1$ term of the interaction dresses quarks of flavor $f$ with condensates of the same flavor,
while $G_2$ leads to mixing.
In particular in the standard NJL Lagrangian  \Eq{eq:NJL}, which is recovered for $G_1 = G_2 = G/2$,
the condensates of both flavors contribute equally to both masses, so that for $m_u=m_d$, the effective mass operators $\hat M_u$ and $\hat M_d$ are always equal as well. 
In this sense \Eq{eq:NJL} corresponds to the maximally flavor-mixed case. 

For $G_2 = 0$, on the other hand, the flavors decouple. 
In this case the quark part of the thermodynamic potential becomes $\Omega_q = \Omega_u + \Omega_d$, 
where $\Omega_f=\Omega_f(T,\mu_f)$ only depends on the chemical potential $\mu_f$. 
The equation of state at nonzero $\mu_Q$ as a function of $\mu$ is then trivially obtained from the one at $\mu_Q=0$
through \Eq{eq:muudQ}.
In particular the phase boundaries related to the properties of the up and down quarks are shifted relative to each other
by $\mu_Q$, so that there are now regions where one flavor is homogeneous while the other one is inhomogeneous.

In ref.~\cite{Frank:2003ve}
a realistic value for the ratio $G_2/G_1$ was estimated to be around $0.2$,
\ie the flavors are neither decoupled nor maximally mixed. 
A detailed discussion of this case will be given in ref.~\cite{dno}.

Finally we note that instead of considering a homogeneous leptonic background and imposing the neutrality condition
for the spatially averaged charge density, we could in principle do better and allow for an inhomogeneous lepton density
as well. 
While the density gradient would increase the kinetic energy of the electrons, it could lower the local Coulomb energy, 
and the optimal shape would result from balancing these two contributions.
This possibility has not yet been considered.
However, given the large incompressibility of electrons, a considerable variation of their density over the length 
scale of the quark structures is unlikely, so that the assumption of a constant electron background is probably not too 
bad.\footnote{We thank the referee for pointing this out to us.}

\subsection{Magnetic fields}
\label{sec:magfields}

A very important element characterizing compact stars is the presence of strong magnetic fields. For magnetars, surface fields 
up to $10^{15} $G have been measured, 
and even higher values are expected to be reached in their cores, possibly reaching up to $10^{18} $G \cite{Lai:1991}. More recently even higher values have been suggested \cite{Ferrer:2010wz}, although under such extreme regimes large deformations are expected to be present and the star structure is likely destabilized \cite{Bocquet,Cardall}.

Aside from the well known effects on the usual homogeneous chiral symmetry breaking mechanism, such as the phenomenon of magnetic catalysis \cite{suga1,Gusyin:1994,Gusynin:1995nb} 
(see \cite{Miransky:2015ava,Andersen:2014xxa} for recent reviews), the presence of strong magnetic fields has a significant influence on the formation and the properties of inhomogeneous phases. 
To see this, we start once again from the simple two-flavor NJL Lagrangian of \Eq{eq:NJL}, and include the coupling with a background magnetic field by the replacement 
$\partial_\mu \, \rightarrow \, D_\mu = \partial_\mu + i {\hat Q} A_\mu$, where $ A_\mu $ is the
electromagnetic field 
and $ {\hat Q} ={\text{ diag}} (e_u, e_d) = {\text{ diag}} (\frac{2}{3} e, -\frac{1}{3} e)$ the electric charge matrix in flavor space ($e$ being the unit electric charge). Due to their unequal charges, the interaction with the magnetic field distinguishes between quark flavors, effectively breaking isospin symmetry.

For a CDW-type ansatz (\Eq{eq:CDW}) in the presence of a static background magnetic field $H$ pointing in the $z$ direction,  the quark energies for each flavor have been calculated in \cite{Frolov:2010} and are given by 
\beq
E_n^f = \pm
\begin{cases}
\sqrt{\Delta^2+p_z^2}+q,&\quad  \, n=0  \\
 \sqrt{\left(\pm\sqrt{\Delta^2+p_z^2}+q\right)^2+2\vert e_f H\vert n} & \quad  \, n>0 \,.
 \end{cases}
 \label{eq:spectrumB}
 \eeq
By inspecting this expression one can see that the quark momenta transverse with respect to the direction of $H$ become quantized into $p_\perp^2 \sim 2 n \vert e_f H\vert$, with the integer $n$ labelling the so-called Landau levels.  In turn, this implies that at the lowest Landau level (LLL), for which $n = 0$, the problem becomes effectively 1+1-dimensional,\footnote{Indeed, the $n=0$ eigenvalues are exactly the energies of a quark in a CDW-type condensate background in the  
chiral Gross-Neveu model.}
a configuration in which inhomogeneous chiral condensation is always favored \cite{Schon:2000he,Basar:2009fg}. 
 
\begin{figure}
\resizebox{0.47\textwidth}{!}{%
  \includegraphics{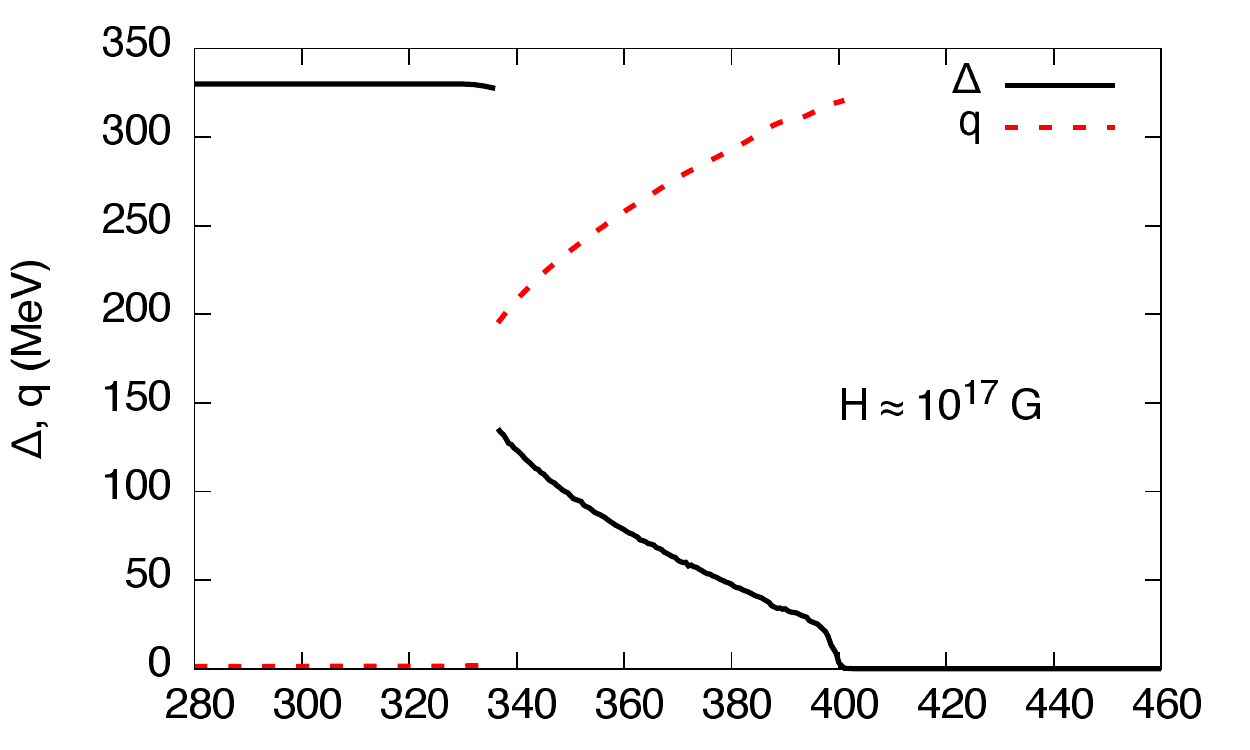}
}
\resizebox{0.47\textwidth}{!}{%
   \includegraphics{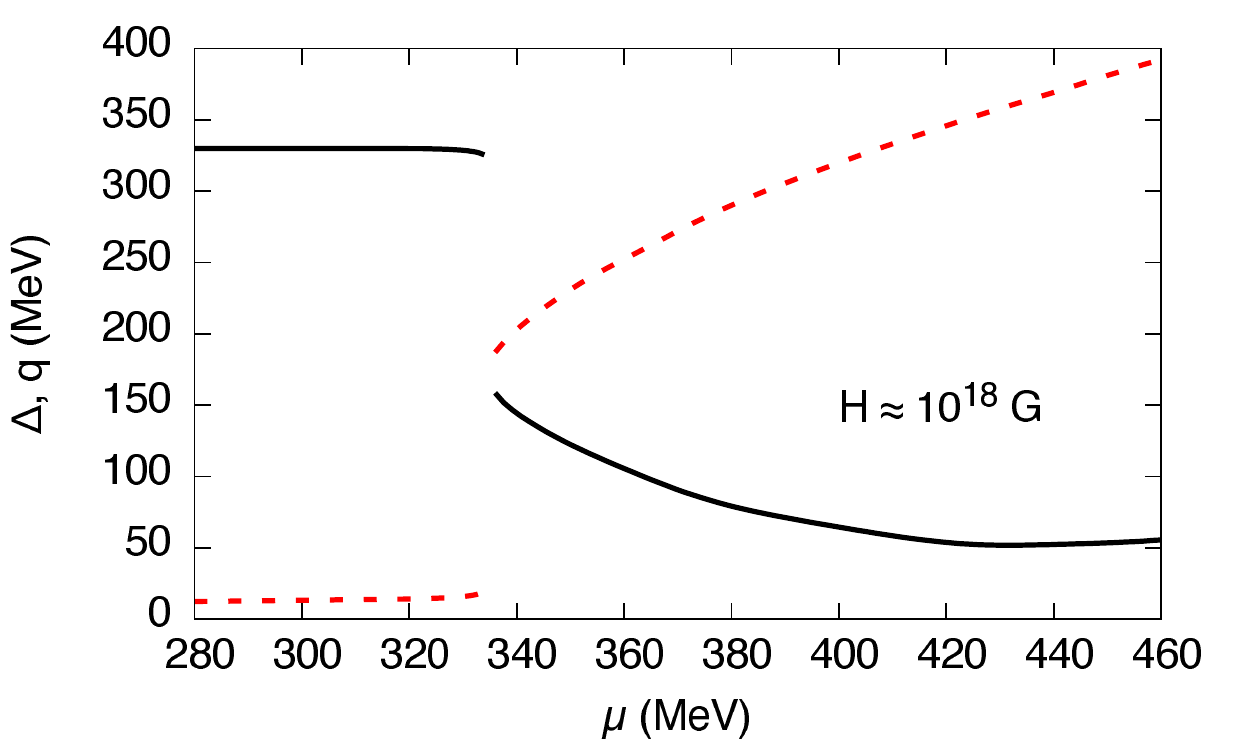}
}
\caption{Order parameters in a magnetic field for a CDW modulation as a function of the quark chemical potential, at given magnetic fields intensities. 
Top: $eH = (28~{\text {MeV}})^2$   ($H = 4.4\times 10^{17} G$),  
bottom: $eH = (90~{\text{MeV}})^2$ ($H = 4.5 \times 10^{18} G$).
Similar results have been presented in ref.~\cite{Frolov:2010}.
}
\label{fig:Bmq}   
\end{figure}

In particular, as a consequence of the spectral asymmetry at the LLL,\footnote{For a more detailed discussion, see \cite{tatplb15}.
} at low temperatures the CDW becomes the thermodynamically favored state over homogeneous matter for any $\mu > 0$.  As can be seen from \Fig{fig:Bmq}, 
at low $\mu$ the system is characterized by a small but nonzero value of the wave number $q$, until a phase transition is reached in the chemical potential region where the inhomogeneous phase would appear even for $H=0$. In the low $\mu$ region the value of $q$ is strongly influenced by the magnetic field, mainly due to the presence of an anomalous term $\sim |eH| \mu q$ in the thermodynamic potential. Therefore the effect is more visible in the lower panel of \Fig{fig:Bmq}, corresponding to a larger value of $H$, 
although it also exist for smaller magnetic fields (upper panel).
After
 the phase transition, the value of $q$ becomes comparable with the  one obtained in absence of a magnetic field, suggesting that the dominant physical mechanism there is the usual particle-hole pairing at the Fermi surface. 
   
If the magnetic field is sufficiently large, the amplitude of the CDW does not melt completely as $\mu$ increases and the inhomogeneous phase extends to higher densities (for a more detailed discussion on the behavior of the CDW order parameters in a magnetic-field background we refer the reader to \cite{Frolov:2010}).  This suggests that chiral restoration might never be reached, although the enhancement of inhomogeneous phases at high densities might 
at least partly be a model artifact~\cite{CB:2011}.
 At high densities we expect in any case color superconductivity to be the favored state for quark matter, see \Sec{sec:CSC}.
 
\begin{figure}
\resizebox{0.47\textwidth}{!}{%
  \includegraphics{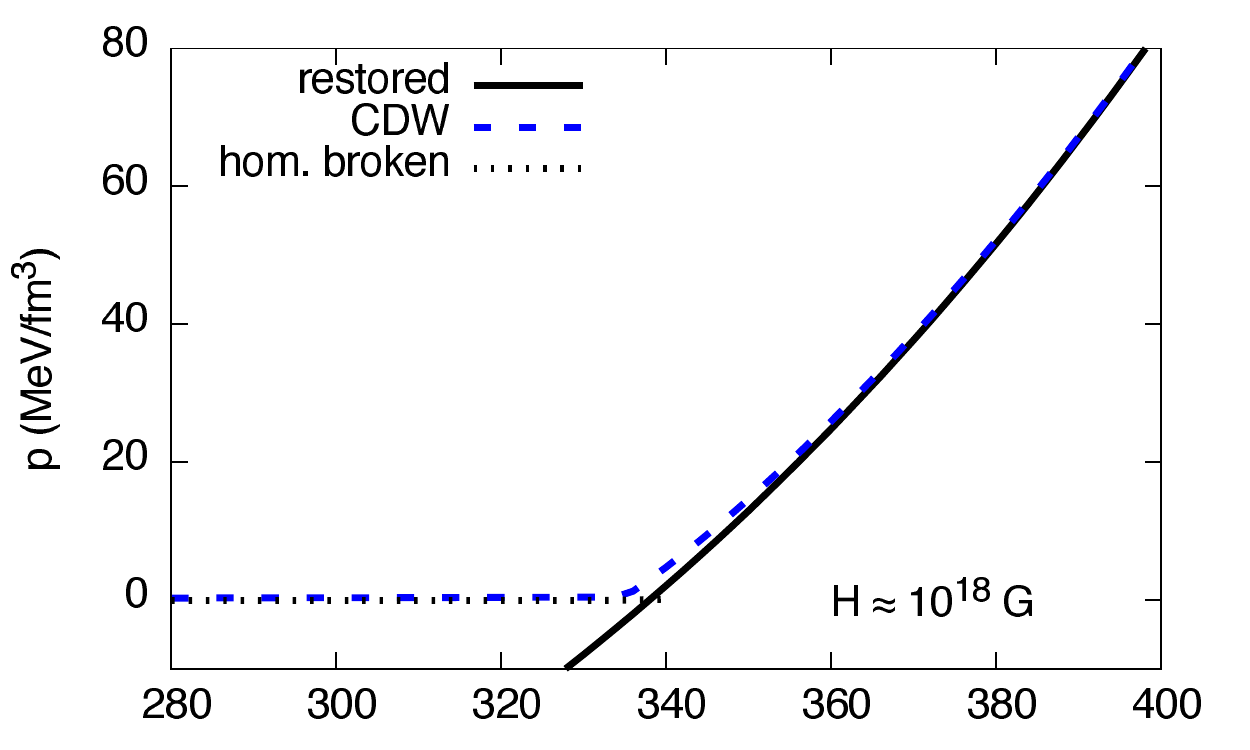} 
  }
  
  \resizebox{0.47\textwidth}{!}{%
  \includegraphics{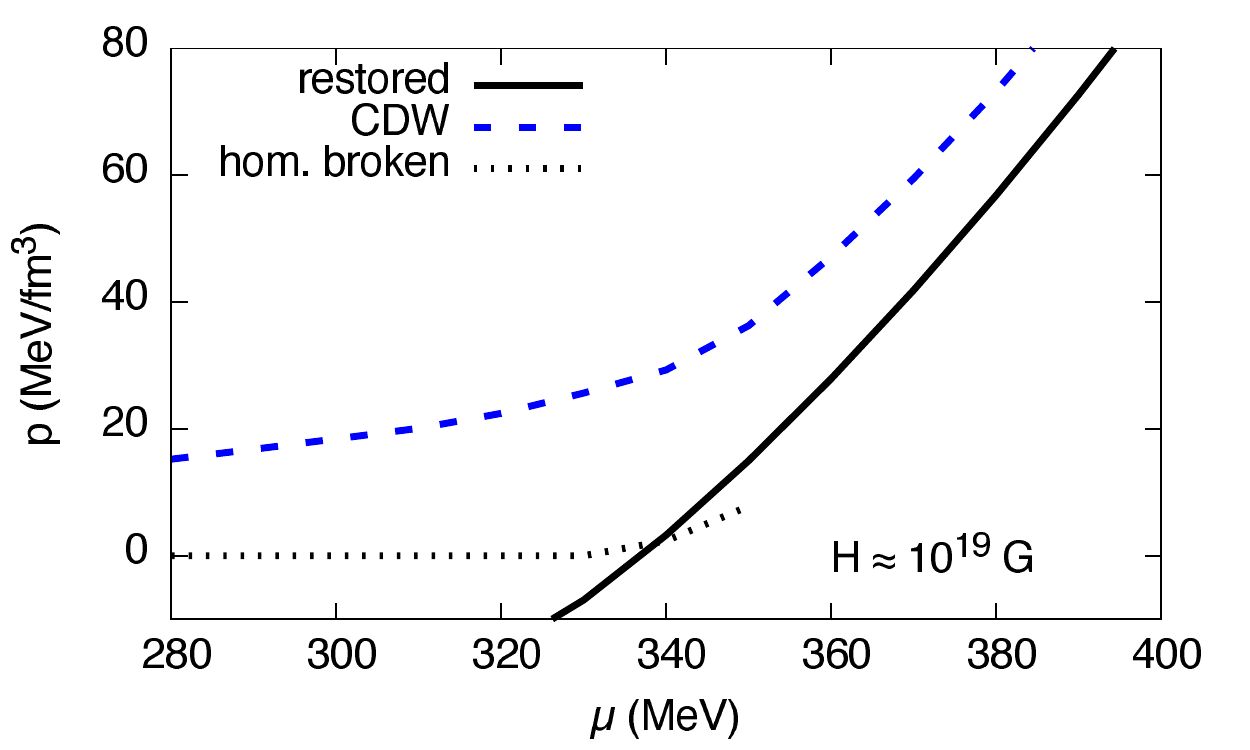} 
  }
  \caption{Pressure at $T=0$ in a background magnetic field
as a function of the quark chemical potential for homogeneous matter and a CDW modulation.
 Top:  $eH = (90~{\text{MeV}})^2$ ($H = 4.5 \times 10^{18} G$)
 bottom:  $eH = (280~{\text{MeV}})^2$ ($H = 4.4 \times 10^{19} G$)
 }
\label{fig:pmag}  
\end{figure}

  Being energetically favored, the CDW phase is characterized by a higher pressure when compared to the homogeneous broken and restored ones, as can be seen from \Fig{fig:pmag}. We note at any rate that this effect is significant only for relatively large values of the magnetic field, while it becomes almost negligible for $H \lesssim 10^{17} G$.

\begin{figure}
\resizebox{0.47\textwidth}{!}{%
  \includegraphics{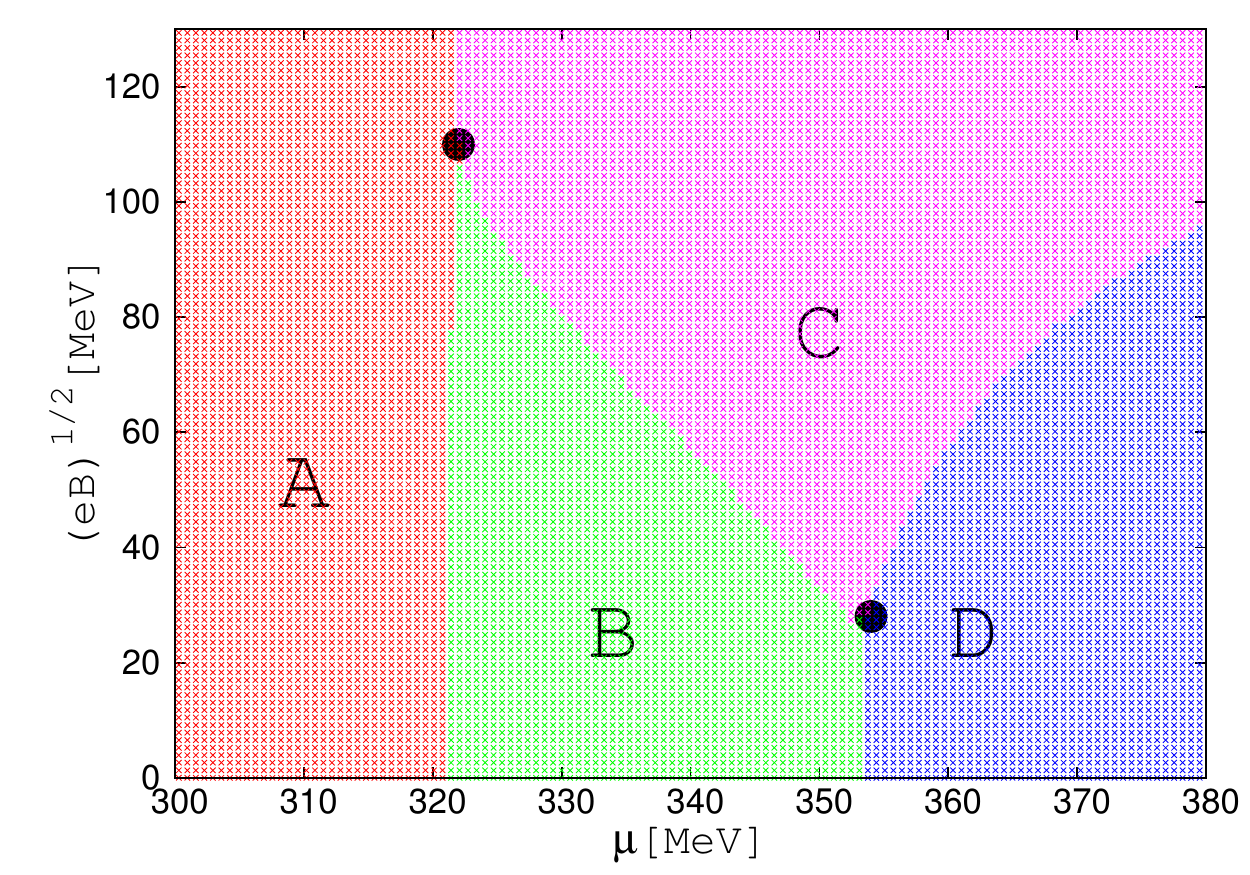} 
  }
  \caption{$\mu-H$ phase diagram allowing for a hybrid chiral condensate. Area A
  corresponds to the CDW phase with $q \propto H$, area B is the HCC phase, 
  area C is the large-$q$ CDW phase and 
  area D is the chirally restored phase.  From ref. \cite{Nishiyama:2015fba}.
  }
\label{fig:hcc}  
\end{figure}

While so far we found that in 3+1 dimensions the RKC is favored over the CDW (cf.~\Fig{fig:pressure}), 
the introduction of a magnetic field and the related dimensional reduction at the LLL might change this picture.
In particular, since in 1+1 dimensional systems complex modulations are typically favored over real ones~\cite{Basar:2009fg}, 
it is possible that the CDW reappears in the phase diagram when a magnetic field is present. 
Moreover, it has been suggested in \cite{Nishiyama:2015fba} that for finite $H$ a hybrid chiral condensate (HCC) akin to the twisted kink crystal introduced in \cite{Basar:2009fg} might become thermodynamically favored. This particular modulation can be interpreted as a generalization of the RKC to include a phase factor, or equivalently as a generalization of the CDW to include a spatially varying amplitude, and is given by
  \beq
  M_{HCC}(z) = \Delta \sqrt{\nu} \mathrm{sn}(\Delta z | \nu)   e^{i q z} \,.
  \eeq
For $q=0$ this solution obviously reduces to the RKC, and if $\nu = 1$ it becomes thermodynamically degenerate with the CDW. If both $q \rightarrow 0$ and $\nu \rightarrow 1$, it approaches the homogeneous limit. While in light of the results discussed in \Sec{sec:model}  we would expect the imaginary part of this modulation to be zero, now the magnetic field always favors nonzero values for $q$, so that the order parameter is always complex. This implies that in the region of the phase diagram where the RKC inhomogeneous island used to appear, the favored solution is now given by the HCC, while in the other regions this ansatz reduces to the CDW modulation, except when chiral symmetry is restored. The $\mu-H$ phase diagram at zero temperature looks then like \Fig{fig:hcc}: for nonzero magnetic fields the low-density region is occupied by the CDW phase with small $q$, which then turns into the HCC after a critical chemical potential is reached. The region of large $\mu$ and low $H$ is chirally restored, while at large magnetic fields and chemical potentials the CDW is again favored, this time with large wave numbers \cite{Nishiyama:2015fba}.

In summary, the presence of a magnetic field qualitatively alters the model phase structure at finite density, dramatically enhancing the size of the inhomogeneous phase. This in turn implies that if quark matter is present in the core of compact stars, it might be entirely in a crystalline state. Even if for small magnetic fields this would only lead to small differences in the equation of state, other quantities like transport properties and neutrino emissivity might be strongly affected, as we will argue in \Sec{sec:implications}.
 
\subsection{Vector interactions}

Vector interactions are known to play an important role in dense matter.
A classical example is the Walecka model~\cite{Walecka:1974qa}, where the repulsive coupling of 
the nucleons to the omega-meson field is crucial 
to reproduce the correct binding energy of nuclear matter at saturation density. 
In compact stars, where matter is expected  to reach several times nuclear-matter density,
vector-interaction effects should therefore be particularly important.

Vector interactions are known to have a strong influence on the properties of homogeneous matter in the NJL model,
where  they cause a significant stiffening of the equation of state and reduce the strength of the first-order phase transition, possibly resulting in the disappearance of the critical point from the phase diagram.

The effect of vector interactions on inhomogeneous matter has been investigated in ref.~\cite{CNB:2010}.
To this end the NJL Lagrangian (\Eq{eq:NJL}) was extended by the term
\beq
\mathcal{L}_V =-G_V (\bar\psi \gamma^\mu \psi)^2 \,,
\label{eq:LGv}
\eeq
where $G_V$ is the vector coupling constant.
Its natural size, obtained from the Fierz transformation of a color-current interaction (``heavy gluon exchange'') is 
one half of the scalar coupling $G$, but fits to the vacuum mass of the omega meson~\cite{Vogl:1991qt,Ebert:1985kz} 
or to finite-temperature lattice-QCD results at vanishing chemical potential~\cite{Steinheimer:2010sp,Steinheimer:2014kka} 
suggest higher or lower values, respectively.
In \cite{CNB:2010} $G_V$ was therefore treated as a free parameter and varied between zero and $G$.

Within the mean-field approximation, only the temporal part 
$V^0 = \langle \bar\psi\gamma^0\psi \rangle$
of the vector condensate, which is identical to the quark number density $n$, was retained.\footnote{
In homogeneous matter the spatial part $\mathbf{V} = \langle \bar\psi \boldsymbol{\gamma}\psi \rangle$ of the vector condensate
does not form, since it would break rotational invariance. 
This argument does not apply to inhomogeneous matter, where rotational symmetry is already broken by the 
scalar and pseudoscalar condensates.
Yet, $\mathbf{V}$ has also a preferred orientation, \ie it breaks parity,
and, at least for the RKC, it  does not appear~\cite{Marco:MSc}.}
The effect of this field can then be absorbed in a shift of the chemical potential,
\beq
\label{eq:mutilde}
\mu \rightarrow \tilde\mu = \mu - 2 G_V n \,,
\eeq
which is a standard technique for homogeneous matter~\cite{Asakawa:1989}.
When dealing with inhomogeneous condensates, the quark density 
is in general
 spatially modulated, implying that the shifted chemical potential also depends on the position. 
 For a CDW this is not the case, as the density is constant, whereas it is for the RKC modulation
(see \Sec{sec:model}). Nevertheless, in \cite{CNB:2010} the shift in the chemical potential  was implemented
only through the spatial average of the vector condensate, an approximation which is expected to be rather inaccurate close to the onset of the inhomogeneous island (cf.~\Fig{fig:RKCshapes}), but otherwise reasonably valid, particularly when approaching the restored phase.

For $G_V=0$, as seen in \Fig{fig:pd},      
the inhomogeneous phase covers the first-order phase boundary between the homogeneous broken and restored phases,
and ends at the critical point~\cite{Nickel:2009wj,Nickel:2009ke}.
Since vector interactions weaken or even remove this homogeneous first-order phase transition,
it was  expected  that the inhomogeneous phase becomes smaller or disappears as well  when $G_V$ is increased.
It turned out, however, that the inhomogeneous phase becomes larger instead,
keeping its extension in temperature and enhancing its size in the chemical potential direction~\cite{CNB:2010}.

\begin{figure}
\resizebox{0.47\textwidth}{!}{%
   \includegraphics{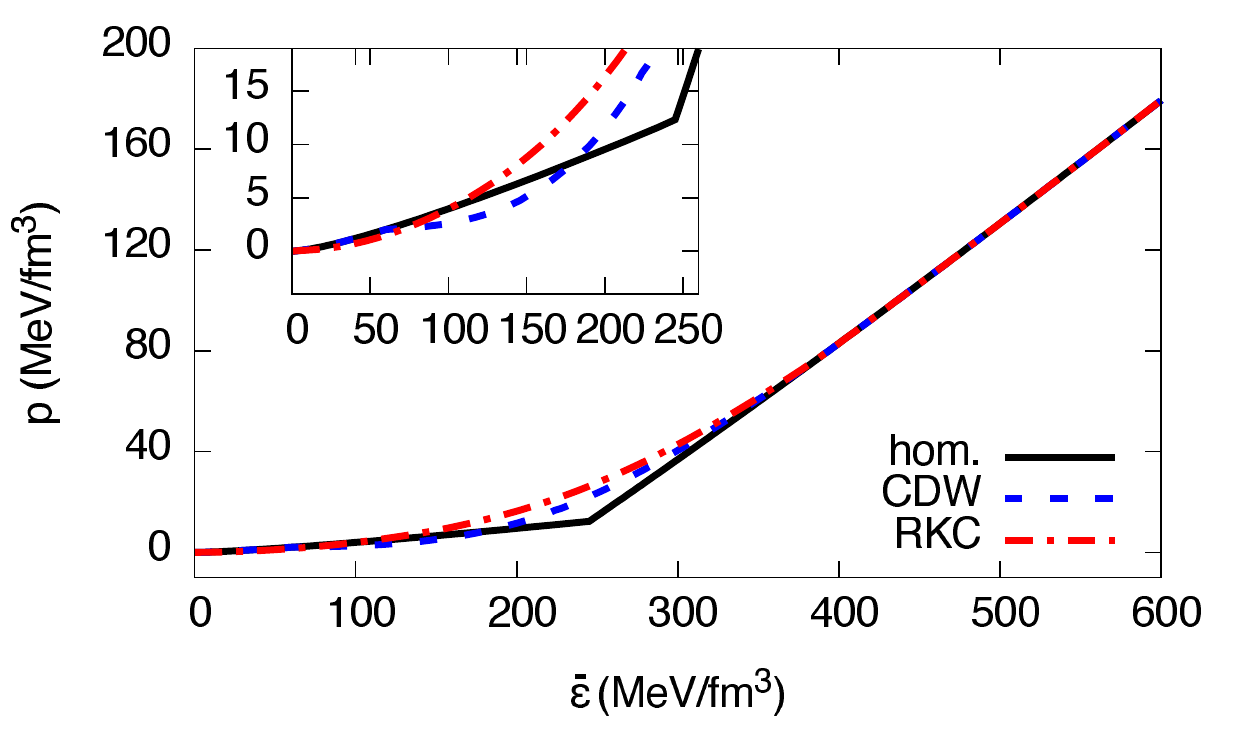}
}
\caption{Equations of state for homogeneous and inhomogeneous (CDW and RKC) quark matter
in the presence of vector interactions with $G_V = G/2$. The inset shows an enlarged detail.
}
\label{fig:vecEoS1}      
\end{figure}

\begin{figure}
\resizebox{0.47\textwidth}{!}{%
   \includegraphics{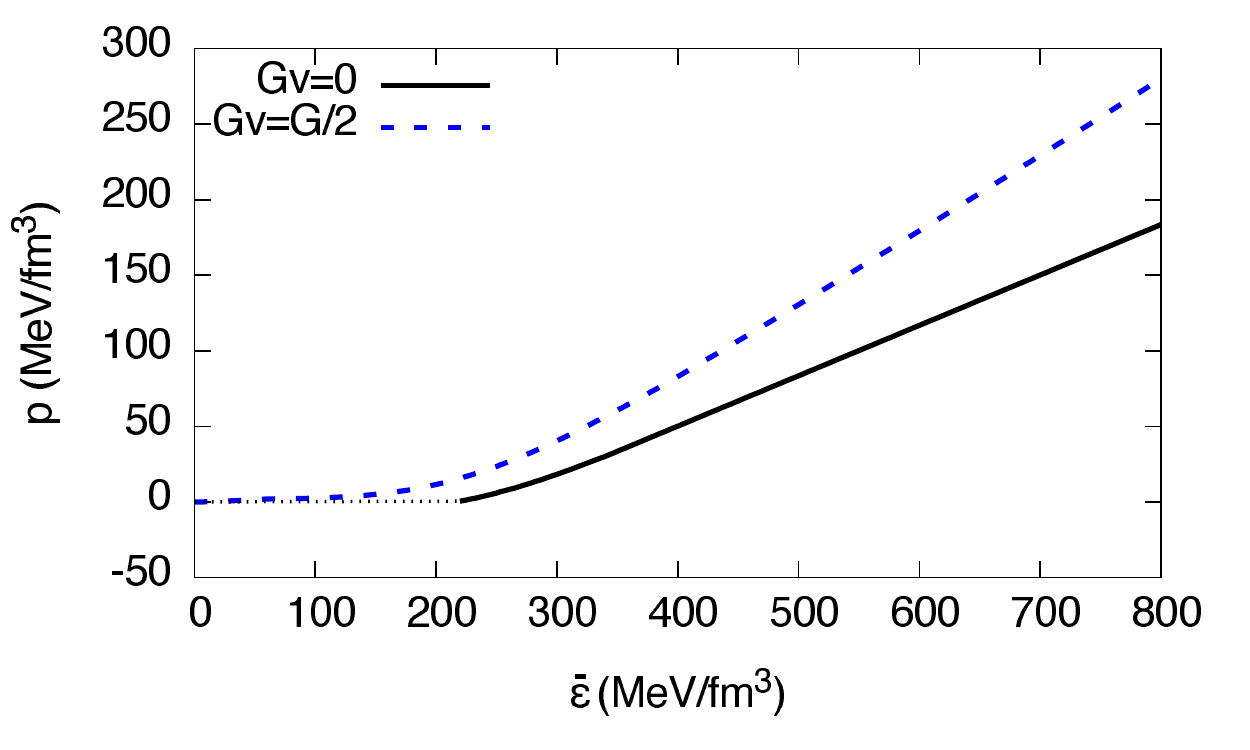}
}

\caption{Equations of state for quark matter with a CDW modulation for $G_V = 0$ and $G_V = G/2$.
}
\label{fig:vecEoS2}      
\end{figure}

In order to investigate the consequences of the vector interaction for the equation of state, 
we show in \Fig{fig:vecEoS1} the results
for homogeneous matter, compared with the RKC and CDW modulations, for $G_V = G/2$. The first visible effect compared to \Fig{fig:eos} is the absence of horizontal jumps in the curves, as all phase transitions become second-order for sufficiently large values of the vector coupling. Additionally, we observe that at low energy densities the equations of state for homogeneous and inhomogeneous matter cross,
\ie inhomogeneous matter, and in particular the RKC, is softer at low and stiffer at high densities. 
Just like for homogeneous matter, the most pronounced effect of including vector interactions is in any case the overall stiffening of the equation of state, as can be seen in \Fig{fig:vecEoS2}, where the pressure with and without vector interactions is plotted for the CDW modulation.

For asymmetric quark matter,  
the isospin structure of the vector interaction becomes again important.
In this context, instead of the pure isoscalar interaction \Eq{eq:LGv}, some authors employ the Lagrangian 
 \cite{Zhang:2009mk,Abuki:2009ba}
\begin{align}
\mathcal{L}_\mathit{{VA}} =-G_V \Huge[ &(\bar\psi \gamma^\mu \psi)^2 + (\bar\psi \gamma^\mu \tau^a\psi)^2 
\nonumber\\
 + &(\bar\psi\gamma^\mu\gamma^5\psi)^2 + (\bar\psi \gamma^\mu\gamma^5 \tau^a \psi)^2 \Huge] \,,
\label{eq:Lviso}
\end{align}
which also contains isovector terms and can be viewed as the vector - axial-vector analogue of the $G_1$ term in 
\Eq{eq:LagNJLiso}.
Indeed, as in the latter,  the different flavor contributions decouple. 
Hence, in
the mean-field approximation 
the effect of the vector part (upper line) reduces 
to a shift of the individual quark chemical potentials for each flavor depending on their own number density, \ie $ \widetilde{\mu}_f = \mu_f -4G_V n_f $.  For two-flavor isospin symmetric matter, where $\mu_u = \mu_d = \mu$ and $n_u = n_d = n/2 $, this reduces to the standard expression \Eq{eq:mutilde}.

Finally we note that, in contrast to \Eq{eq:LGv},  $\mathcal{L}_\mathit{VA}$ contains axial-vector terms. 
These are required in the isovector channel by chiral symmetry. 
Whether the corresponding mean fields  have an effect, \eg on a CDW, has not yet been investigated.

\subsection{Competition with color superconductivity}
\label{sec:CSC}

So far we have concentrated on homogeneous and inhomogeneous chiral condensates, 
\ie the condensation of quarks with antiquarks or quark-holes. 
It is known however that at very high densities particle-hole pairing is disfavored against color superconductivity (CSC)
\cite{Shuster}, where quarks are paired with other quarks.
Moreover, NJL-model \cite{Berges:1998rc,Schwarz:1999dj,Buballa:2005}
and Dyson-Schwinger \cite{Muller:2013pya} studies of isospin-sym\-metric homogeneous matter suggest 
that CSC phases appear already at the chiral restoration transition.
This raises the question of how the inhomogeneous chiral condensates, which are found in the same region,
are affected by CSC, and in particular whether they survive at all.

In refs.~\cite{Sadzikowski:2002iy,Sadzikowski:2006jq} this was investigated within a two-flavor NJL model.
To this end, the Lagrangian \Eq{eq:NJL} was augmented by the term
\beq
      \mathcal{L}_D = G_D \left( \psi^T C \gamma^5 \tau^2 \lambda^A\psi\right)^\dagger
                                         \left( \psi^T C \gamma^5 \tau^2 \lambda^A\psi\right) ,
  \label{eq:LagD}                                       
\eeq
where $\lambda^A$, $A=2,5,7$ denote the antisymmetric Gell-Mann matrices in color space
and $C$ is the charge conjugation matrix.
The coupling constant $G_D$ is not strongly constrained. Similar to the vector coupling $G_V$, it is often treated 
as an additional parameter, typically of the order of $G$ or less\footnote{Performing a Fierz tranformation of a color-current
interaction yields $G_D = 0.75\,G$.},
although slightly larger values have also been employed in compact-star phenomenology~\cite{Klahn:2006iw}.

Eq.~(\ref{eq:LagD}) describes a quark-quark interaction in
the spin-0 color-antitriplet channel, which is expected to be the most attractive one.
The corresponding diquark condensate
\beq
       \langle \psi^T C \gamma^5 \tau^2 \lambda^A\psi\rangle = \frac{\Delta_D}{2G_D}
\label{eq:2SC}
\eeq
is an order-parameter for the so-called two-flavor superconducting (2SC) phase.
It is proportional to the gap parameter $\Delta_D$, which can be identified with the gap in the quark 
spectrum if no other condensates are present. 

In refs.~\cite{Sadzikowski:2002iy,Sadzikowski:2006jq} 
the formation of a nonzero (albeit spatially constant) diquark gap $\Delta_D$ was studied together with a CDW ansatz (\Eq{eq:CDWcond}) in the chiral sector.
Despite this complication, the author was still able to find the eigenvalues of the effective Dirac Hamiltonian
analytically, keeping the combined mean-field problem on a manageable level.
The phase structure was then obtained by minimizing the thermodynamic potential with respect to 
the amplitude $\Delta$ and wave number $q$ of the CDW, together with the gap parameter $\Delta_D$.
  
 \begin{figure}
\resizebox{0.47\textwidth}{!}{%
  \includegraphics{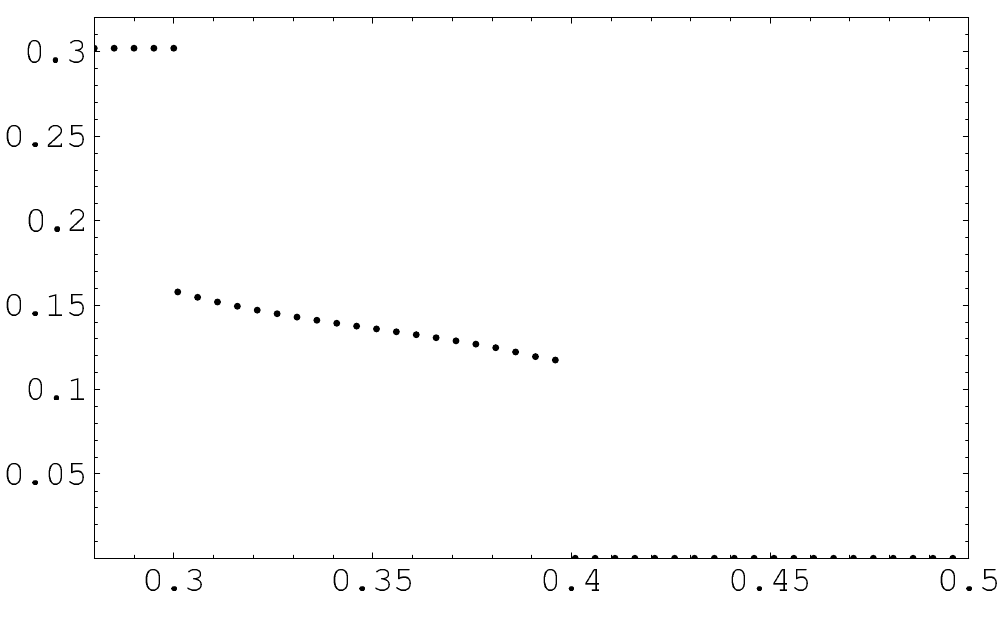}
}

\resizebox{0.47\textwidth}{!}{%
  \includegraphics{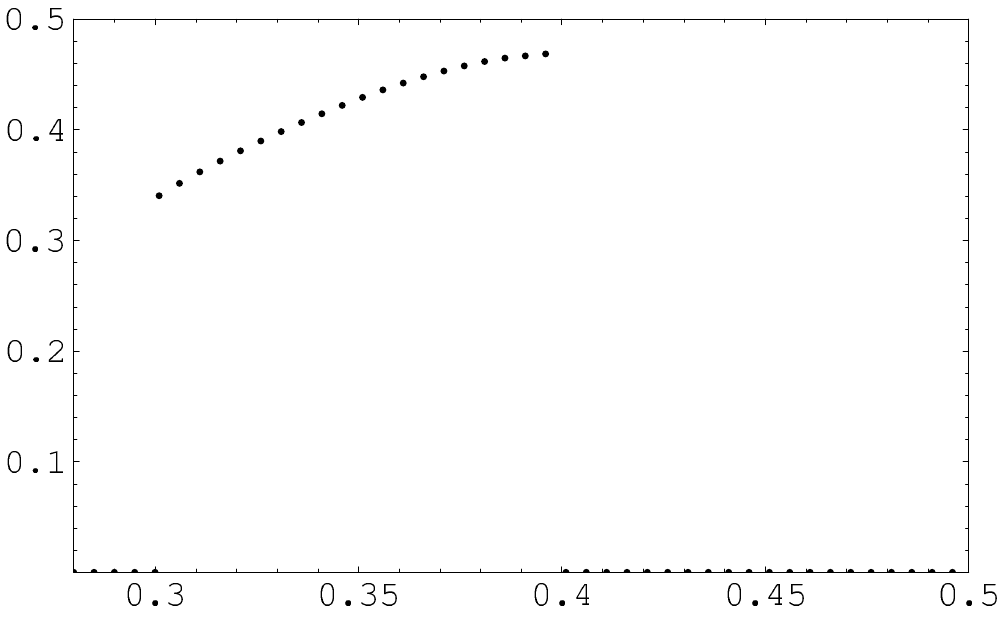}
}

\resizebox{0.47\textwidth}{!}{%
  \includegraphics{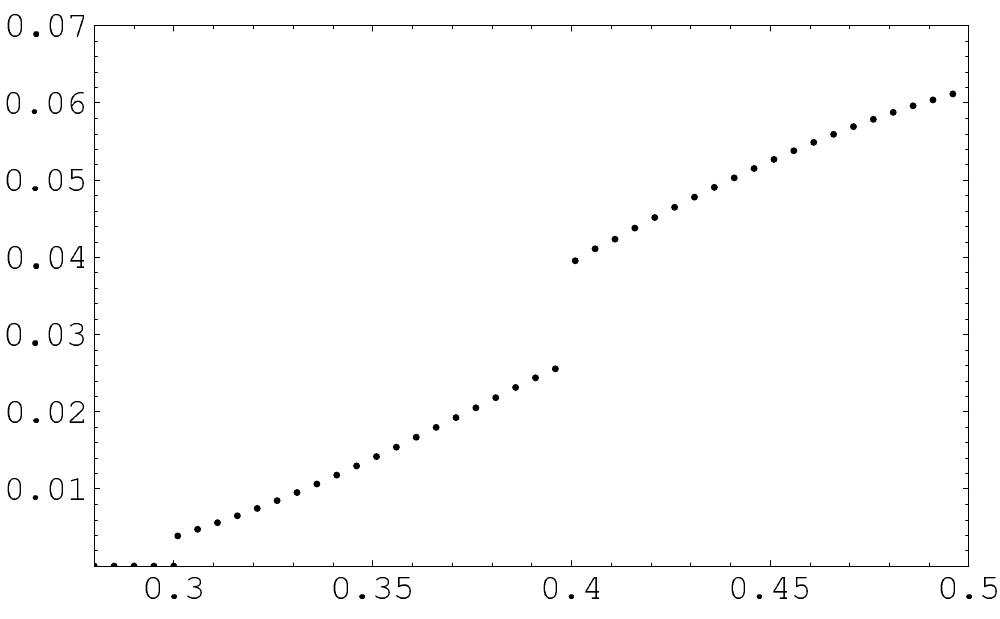}
}
\caption{Amplitude $\Delta$ (top) and wave number $q$ (middle) of the CDW modulation, 
and 2SC gap parameter $\Delta_D$ (bottom) for $G_D = G/2$  as functions of $\mu$ at $T=0$.
All numbers are given in units of GeV.
From~\cite{Sadzikowski:2002iy}.
}
\label{fig:csc}      
\end{figure}

For $G_D = G/2$ the results at $T=0$ as functions of the chemical potential are displayed in \Fig{fig:csc}.
One can see three different phases, separated by first-order phase transitions: 
For $\mu < 300$~MeV, we have $\Delta \neq 0$ and $q = \Delta_D = 0$, \ie 
the normal (non-superconducting) homogeneous chirally broken phase is favored. 
For $\mu > 400$~MeV, on the other hand, we have $\Delta_D \neq 0$ and $\Delta = 0$,
corresponding to the 2SC phase with restored chiral symmetry. 
Finally, the most interesting regime is the intermediate one, where the CDW ($\Delta$, $q \neq 0$)
coexists with the CSC gap ($\Delta_D \neq 0$). 
Note that there is no region where the CDW exists without CSC.
This can be traced back to the fact that at $T=0$ the Cooper instability which leads to $\Delta_D \neq 0$ 
occurs as soon as there is a nonzero density. 
At larger temperatures, on the other hand, CDW regions without CSC are possible~\cite{Sadzikowski:2006jq}.

From the above argument, it also follows that for small values of $G_D$ the location of the coexistence region at $T=0$
should roughly coincide with the inhomogeneous phase in the case without considering CSC.\footnote{Note that 
a different regularization was used in ref.~\cite{Sadzikowski:2002iy}, so that we cannot directly compare it with 
the other results we have shown.}
Increasing $G_D$ favors phases with large diquark gaps. 
As a consequence all phase boundaries move towards lower chemical potentials, albeit in an asymmetric way,
so that the size of the coexistence phase decreases~\cite{Sadzikowski:2002iy}.

It is important to note that these studies on the interplay of inhomogeneous chiral symmetry breaking with color superconductivity
 have been performed for isospin symmetric matter only, and that the picture described here might be significantly altered if this condition is relaxed. 
Indeed, it is known that imposing electric neutrality strongly disfavors 2SC pairing, since the Fermi surfaces of the up and down quarks,
which are paired in the condensate (\ref{eq:2SC}), are pulled apart \cite{Alford:2002kj}.
In particular, the homogeneous 2SC phase could become unstable with respect to the formation of inhomogeneous 
color-superconducting phases~\cite{Giannakis:2004pf}. 
Since the corresponding free-energy gains are much smaller than for homogeneous BCS pairing in isospin symmetric 
matter, there should be more room for inhomogeneous chiral condensates, which, as we have seen, 
are much less sensitive to isospin asymmetry.

\subsection{Strange quarks}

If quarks are present in the cores of compact stars, the density might even be high enough to populate strange-quark states.
This is a rather common scenario in bag models where the strange quark mass is usually taken to be of the order of its 
bare value, $\lesssim 150$~MeV, well below the associated chemical potential.   
Since strange quarks are negatively charged, they help neutralizing the matter and thus reduce the isospin asymmetry. 
Moreover, a high abundance of strange quarks
favors color superconductivity in the color-flavor locked (CFL) phase
where up-, down and strange quarks are paired in a symmetric way. 
This phase is electrically neutral even in absence of electrons, \ie for $\mu_Q=0$~\cite{Rajagopal:2000ff}.

In the NJL-model, on the other hand, strange quarks have considerably higher dynamically masses and therefore
only appear at relatively high chemical potentials,  $\mu \gtrsim 400$~MeV~\cite{Ruester:2005jc}
or even 500~MeV \cite{Abuki:2004zk}.\footnote{Similar thresholds have also been found in Dyson-Schwinger QCD studies of
color-superconducting matter without neutrality constraints~\cite{Muller:2013pya}.} 
As a consequence, 
it is typically found that the densities needed to have sizeable amounts of 
strange quarks (and in particular CFL pairing) are either not reached in compact stars, or that 
the softening of the equation of state due to the appearance 
of strange quarks immediately renders the star unstable~\cite{Klahn:2006iw,Baldo:2002ju,Buballa:2003et}. 

Nevertheless, even if their density vanishes, strange-quark degrees of freedom can influence the equation of state via
flavor mixing with the up and down quarks.  
This mixing is caused by the 't Hooft interaction 
\beq
      \mathcal{L}_H = \kappa \left( {\det}_f \left[\bar\psi \frac{1-\gamma^5}{2}\psi\right] + {\det}_f \left[\bar\psi \frac{1+\gamma^5}{2}\psi\right] \right) ,
  \label{eq:LagH}                                       
\eeq
which corresponds to the three-flavor analogue to the $G_2$-term in \Eq{eq:LagNJLiso}.
Here ${\det}_f$ denotes a determinant in flavor space, i.e., $\mathcal{L}_H$ is a six-point interaction,
mixing all three flavors.
As a consequence the strange quark-condensate, which enters the condensate part of the thermodynamic potential,
is sensitive to the density of the non-strange quarks and therefore varies already below the threshold for strange quarks.

Unfortunately, while there exist several publications on three-flavor crystalline color superconductors,
there are so far only very few works dealing with inhomogeneous chiral condensates in three-flavor matter.
The first and most elaborate one is Ref.~\cite{Moreira:2013ura}, where a three-flavor NJL model with  't Hooft interaction was 
employed.
Within that model the authors considered a generalized CDW ansatz, 
keeping the condensate functions \Eq{eq:CDWcond} in the up- and down-quark sector, 
while the strange-quark condensate was taken as a homogeneous background,  $\langle\bar s s\rangle = \mathit{const}$.
For this ansatz the mean-field Hamiltonian could still be diagonalized analytically, thus keeping the problem on a 
tractable level.
Nevertheless, since the strange and non-strange sectors are coupled through the 't Hooft term,
the strange-quark condensate has a non-trivial effect on the phase structure.  
In particular it was found that the chemical-potential window where the inhomogeneous solution is favored can increase 
considerably.

The calculations of Ref.~\cite{Moreira:2013ura} have been performed for a common quark chemical potential.
It is likely that a similar enhancement of the inhomogeneous phase will also occur when neutrality conditions are imposed, although this still needs to be  checked explicitly.

In Ref.~\cite{Carignano:2015kda} the same CDW-like modulation was considered in electrically neutral matter but without 't Hooft interaction. 
The influence of the strange quarks on the non-strange sector in this case is therefore only indirect via the neutrality condition,
and hence, their effect on the inhomogeneous phase is quite marginal, unless very high densities are reached in the core of compact stars.

\section{Astrophysical implications of inhomogeneous chiral symmetry breaking}
\label{sec:implications}

After describing the thermodynamical properties of inhomogeneous chiral symmetry breaking phases within effective models of cold and dense quark matter, we now want to discuss possible consequences of the formation of these crystalline structures on phenomenological observables related to compact stars. 

The determination of clear signatures relating macroscopic observables with the microscopic properties of matter within 
 stellar objects is obviously a very challenging task. 
From the results for the equations of state we have discussed earlier
we can already anticipate that the effect of inhomogeneous condensates on the mass-radius relation will be marginal.
This will be discussed in \Sec{sec:MR}.
More promising signatures can be expected from observables related to transport properties, 
since they are directly influenced by the crystalline structure of the matter and the corresponding low-energy
excitations (phonons).
Some of these signatures, like gravitational waves and glitches,
have been investigated for crystalline color superconductors (cf.~ref.~\cite{Anglani:2013gfu}),
but not yet for the specific case of inhomogeneous chiral condensates.
An exception is the neutrino emissivity, which has been worked out
for a CDW in ref.~ \cite{Tatsumi:2014cea}.
This will be reviewed in \Sec{sec:nuemi}.

\subsection{Mass-radius relations}
\label{sec:MR}

\begin{figure}
\resizebox{0.47\textwidth}{!}{%
  \includegraphics{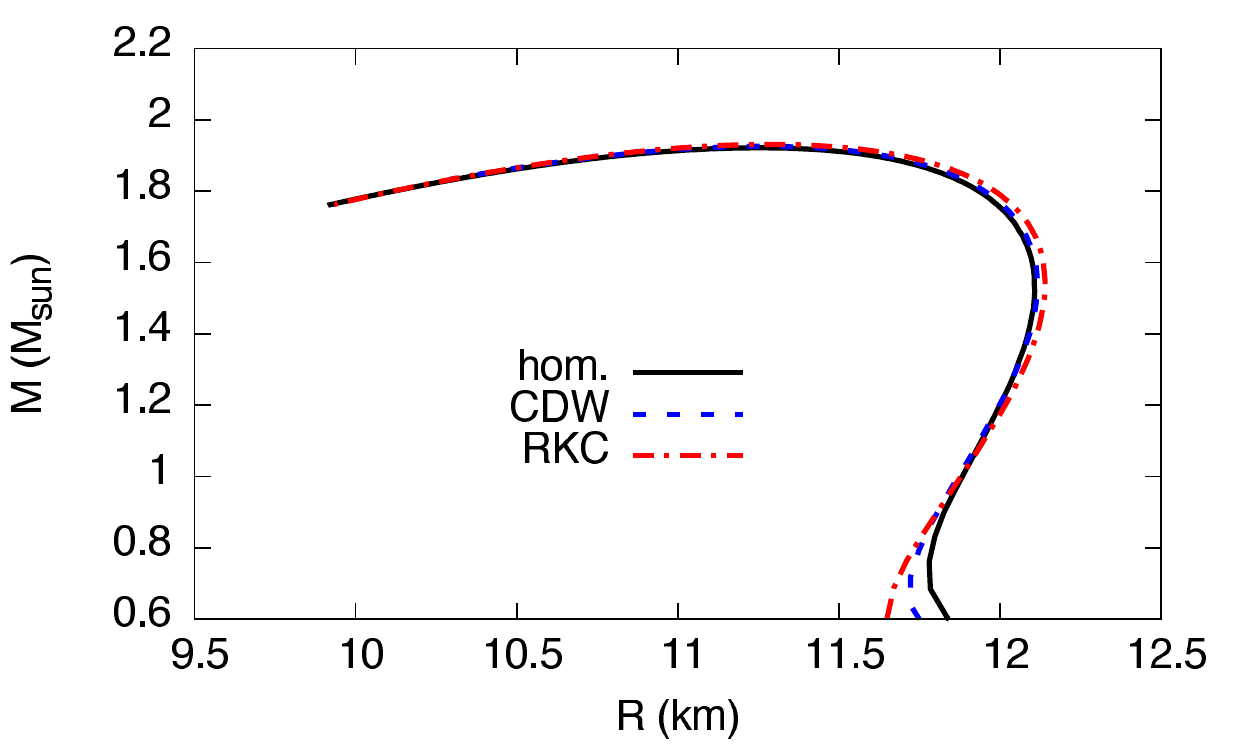}
}

\resizebox{0.48\textwidth}{!}{%
  \includegraphics{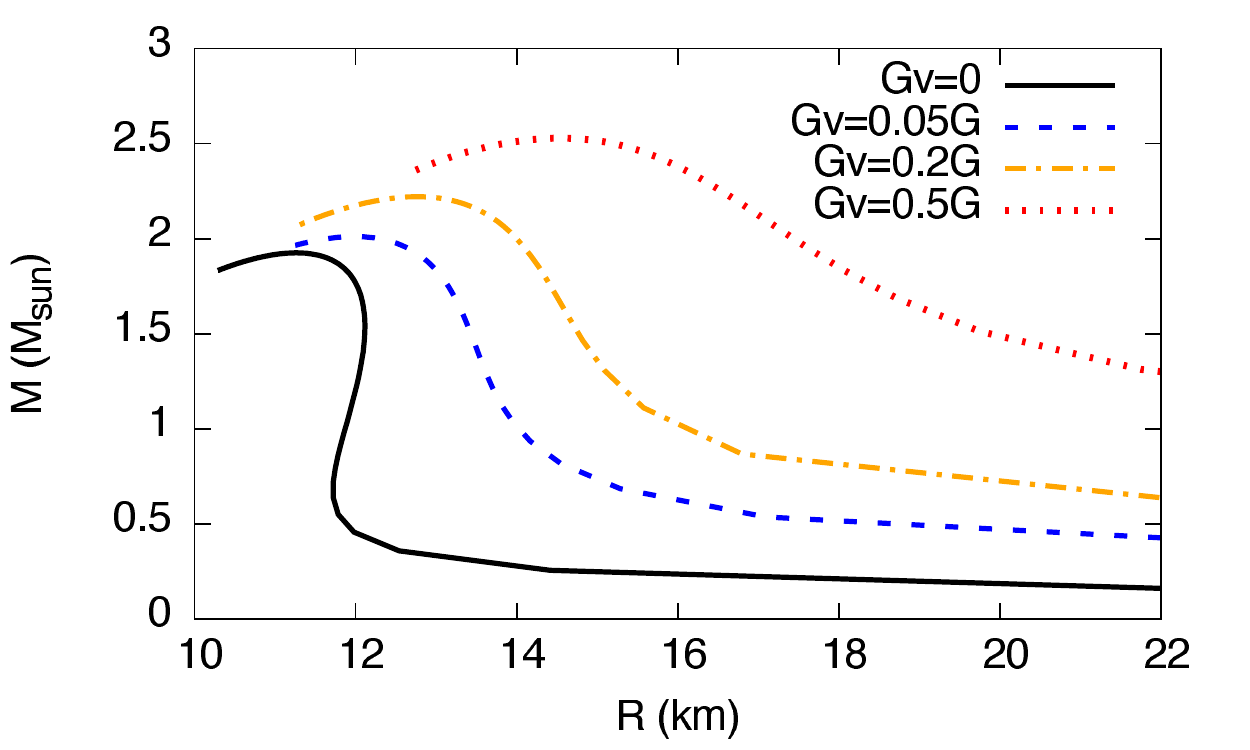}
}
\caption{
Mass-radius sequences for a pure quark star,
neglecting electric neutrality and magnetic fields.
Top: Comparison of three different modulations for $G_V = 0$.
Bottom: CDW modulation with different strengths of the vector coupling.}   
\label{fig:MRquarks}    
\end{figure}

A possible way of discerning between different microscopic compositions of matter in a compact star is to look at the relationship between its mass and radius.  Mass-radius sequences relating these two quantities can be obtained by solving the Tolman-Oppenheimer-Volkoff (TOV) equation \cite{Oppenheimer:1939ne,Tolman:1939jz}, 
which uses the equation of state of the system as an input.
Until relatively recently, this method has been unable to provide clear information, as simultaneous measurements of masses and radii of compact stars are extremely challenging and plagued with large uncertainties. 
The situation
partially changed in the past few years thanks to the very precise measurements of pulsars with unusually large masses, close to 2$M_\odot$ \cite{Demorest,Antoniadis}, $M_\odot$ being the solar mass. These measurements allowed to rule out a large number of softer equations of state, which would be unable to support such massive stars, and serve as a benchmark for all possible models describing the microscopic composition of matter inside these compact objects.

As discussed in sects.~\ref{sec:model} and \ref{sec:csp}, the appearance of inhomogeneous condensates results in relatively small variations of the equation of state,
especially when compared with the uncertainties caused by using different models and parametrizations.
As a consequence, the effects of spatial modulations on the mass-radius relations are very small as well.
To see this, we show in \Fig{fig:MRquarks} mass-radius sequences of different equations of state.
Since  at this point, rather than providing a realistic description of stellar matter,
we just want  to compare the influence of different mechanisms, we consider pure quark stars and neglect electric 
neutrality and magnetic fields.
In the upper panel the equations of state of \Fig{fig:eos} have been used, corresponding to 
different spatial modulations in matter without vector interactions.\footnote{
Note that these equations of state do not correspond to self-bound quark matter. 
Therefore the curves turn around towards large radii at low masses, similar to ordinary neutron stars, but unlike
strange stars in the case of absolutely stable strange quark matter.
}
Except for the low-mass region, which should become irrelevant once a more realistic description of the hadronic 
phase is implemented, the effect of the inhomogeneity is negligible.
Comparing this result with the curves shown in the lower panel,
which have all been obtained with the CDW ansatz, but for different values of the vector coupling,
we conclude that it would be impossible to infer the existence of an inhomogeneous phase on the basis of a 
mass-radius measurement, even if it was very precise.

On the other hand we have seen in \Sec{sec:csp} that the inhomogeneous phase itself is rather robust 
and that its size is even increased by the presence of vector interactions, strange-quark degrees of freedom,
and magnetic fields. 
This means that, as long as there is a quark phase at all in compact stars, it is quite likely that it is inhomogeneous.\footnote{
If high enough densities are reached without having a phase transition to quark matter, inhomogeneous 
chiral condensates could also become favored in the hadronic phase~\cite{Heinz:2013hza}. 
}

In order to determine whether a compact star with a crystalline core is compatible with the 2$M_\odot$ measurements, 
the authors of \cite{Carignano:2015kda} investigated solutions of the TOV equation for 
a hybrid configuration, consisting of an inhomogeneous quark-matter core surrounded by nuclear matter.
The latter was described using the non-linear Walecka model, while for the quark phase
the authors considered a CDW modulation within
an extended NJL model. Vector interactions,  electric neutrality and beta equilibrium were implemented, considering for simplicity a Lagrangian without flavor mixing,
\ie  \Eq{eq:LagNJLiso} with $G_2=0$ for the scalar-pseudoscalar interaction and \Eq{eq:Lviso} for the vector interaction. 
The effect of strangeness was also studied, employing three-flavor versions of the NJL 
model (without 't Hooft interaction) and of the non-linear Walecka model.
  Following ref.~\cite{Moreira:2013ura}
inhomogeneous quark matter was then set up as a CDW in the up- and down quark sectors 
in the homogeneous background of a spatially constant strange-quark condensate.
Color supercon\-ductivity, on the other hand, was not considered.
Finally, a magnetic field  with a chemical-potential dependent strength  
varying from $10^{15} $G at the surface to $2.5\times 10^{18} $G in the core  was included in the model.

Within this approach, hybrid stars with maximum mass\-es above 2 $M_\odot$ 
and containing a quark-matter core in the inhomogeneous phase
have been obtained, suggesting that such configurations are possible in nature.
It is important however 
 to recall that in the presence of magnetic fields, neutron stars
are no longer spherical.
 Therefore the use of 
the TOV equations in this case is incorrect, and a calculation which properly takes into account 
the deformed star shape is required \cite{Bocquet,Cardall,Chatterjee:2014}.  
Nevertheless, the expected influence of the magnetic field on the maximum mass 
is relatively small \cite{Bocquet}, so that 
a more refined calculation would likely lead to values which are compatible
with the ones of  \cite{Carignano:2015kda},
 especially if one considers the large uncertainties originating from the use of phenomenological models with free parameters.
A numerical cross-check of this expectation would be of course extremely desirable.

A major uncertainty in modeling hybrid stars arises from the fact  that usually 
two different models are employed to describe hadron and quark phases.
As a consequence there is in principle an unknown parameter (``bag constant'') which determines the relative
difference of the pressure in the two models~\cite{Pagliara:2007ph}. 
In addition, there are different ways how to join them.    
While in most calculations, including Ref.~\cite{Carignano:2015kda},  a Maxwell construction is performed, 
it was argued in Ref.~\cite{Glendenning:1992vb} that the more correct approach is a Gibbs construction.
This is true, however, only for sufficiently small values of the surface tension, 
which is practically unknown, thus being another source of uncertainties. 

Ultimately, one should therefore aim for a unified description of the quark and hadron phases.
A model allowing for inhomogeneous quark matter might be a step in the right direction~\cite{Buballa:2012vm}, as it could allow 
for quarks to lump together into baryons. Indeed, it is known since more than 25 years that baryons in the NJL model 
can be described as non-topological solitons~\cite{Alkofer:1994ph,Christov:1995vm,Ripka}, 
which are associated with a spherically symmetric inhomogeneous chiral
field. Obviously it would be very exciting to pursue this idea further.

\subsection{Neutrino emissivity}
\label{sec:nuemi}
A perhaps more promising way to get insights about
 the composition of compact stars is to observe their cooling process, which is expected to happen mainly through neutrino emission. 
 A correct phenomenological description of this process should be able to account for the relatively low surface temperatures of young pulsars like the Vela or 3C58~\cite{Pavlov:Vela,Slane:3C58},
without conflicting with the slower cooling of Cas A~\cite{Ho:CasA}. 
In particular, 
since the mass of Cas A was estimated to be $\gtrsim 1.5 M_\odot$,
the cooling rate should not rise too fast with the mass of the star. 
This suggests that some exotic cooling scenario might be required to explain these results. 

Among these mechanisms, quark cooling is very efficient, but the direct Urca processes $d \rightarrow u + e^- + \bar\nu_e $ and
$u  + e^- \rightarrow d + \nu_e $ are kinematically suppressed for non-interacting massless quarks. This is encoded in the so-called triangular condition, a direct consequence of energy and momentum conservation stating that in this case
 all momenta must be collinear, thus dramatically reducing the phase space available. This in turn implies that in the standard scenario of a high-density phase of chirally restored quarks, at least for low temperatures, these processes cannot account for the very rapid cooling observed. 
The picture however changes significantly when inhomogeneous phases are included: if a crystalline chiral condensate forms, quark momenta will no longer be conserved due to their interactions with it.  The inhomogeneous condensate therefore provides the additional momentum which might relax the triangular condition and make these processes kinematically allowed, without the need for spectator particles.

This scenario has been worked out in detail in \cite{Tatsumi:2014cea}, where the neutrino emissivity 
(\ie the emitted energy per unit volume and time)
has been calculated for the case of a CDW modulation (\Eq{eq:CDW}) within a simple two-flavor NJL model. In the following we will report the main results of their calculation.
  Since a full computation becomes extremely involved due to the deformation of the Fermi surface generated by the formation of inhomogeneous condensates, the authors focused on two limiting cases: the onset of the inhomogeneous phase, where the Fermi sphere deformation is relatively small, and the region close to the chiral restoration transition, where the deformation is instead very pronounced.  In the first case, the result for the emissivity is given by 
\beq
\epsilon^{\rm onset}_{\rm CDW}\simeq \frac{457}{1680}\pi {\tilde G}_F^2 \mu_u\mu_d\frac{\mu_e^2}{q}T^6 \,,
\label{eq:emisonsetfinal}
\eeq
while in the second it is 
\begin{align}
\epsilon^{\rm rest}_{\rm CDW} & =\frac{1}{2}\frac{457}{1680}\pi \tilde G_F^2 \Delta^2 q\frac{\mu_e}{\mu_u} T^6\cr \\
&\times\left\{\log\left(\frac{2q}{\Delta}\right)+\frac{\mu_d}{\mu_u}\left[\log\left(\frac{2q}{\Delta}\right)-2\right] \right\} \,.
\label{eq:emisfinal}
\end{align}
Here $\tilde{G}_F = G_F \cos \theta_C$, with $G_F$ weak coupling constant and $\theta_C$ Cabibbo angle. 
For the first case, the dependence on the wave number of the CDW modulation is of the form $1/q$, arising from 
the integration over the phase factor $e^{-i q z}$ associated with the inhomogeneous chiral condensate. This specific
momentum dependence is similar to the one obtained for pion condensation, and the resulting magnitude of the emissivity is comparable to the efficient one obtained for massive quarks.   
Close to the restoration transition the $q$-dependence becomes instead more involved, and the emissivity is typically up to two orders of magnitude lower.
At any rate, both results are significantly larger than standard processes like the modified Urca. This allows for quick cooling, similar to 
the Urca process with interacting quarks~ \cite{Iwamoto:1980eb} or
other exotic mechanisms such as the ones related to the presence of a pion condensate \cite{Maxwell:1977zz}.

While this calculation has been performed only for the specific case of a CDW modulation, we can expect neutrino emission to be significantly enhanced for any kind of modulation of the chiral condensate. A similar calculation for the RKC would be desirable, but likely to be more involved due to the more complicated structure of the quark eigenfunctions required to calculate the transition matrix elements.  

Ultimately, the exact cooling rate of a star will of course depend on the extension of the inhomogeneous window. This in turn implies that cooling measurements might in principle impose limitations on the size of inhomogeneous phases: if crystalline chiral condensates form only in a limited range of densities,
cooling will be most efficient within a relatively thin shell inside the star. 
Hence, as long as chiral restoration is reached in the center of the star, 
the emitted energy grows only like the surface of that shell.
The authors of ref.~\cite{Tatsumi:2014cea} therefore conclude that 
heavier stars may not cool faster than lighter ones.

Conversely, if the inhomogeneous phase is very broad, more massive stars will have a larger region occupied by crystalline condensates and will cool more efficiently. 
In order to support these qualitative arguments with a more quantitative estimate, however, more thorough investigations are needed, most notably an evaluation of the heat capacity of the inhomogeneous layer,
as well as its thermal coupling to the other regions of the star.

\section{Conclusions }

We discussed the phenomenon of inhomogeneous chiral symmetry breaking, with a particular focus on its astrophysical implications. 
Several models suggest that
 the formation of spatially modulated structures of the chiral order parameters is energetically favored in a window of the phase diagram which is relevant for cold and dense stellar matter. 
After introducing some basic thermodynamical properties of these crystalline solutions, we
discussed how they are affected by including several necessary constraints for a correct description of matter in compact stars.
While electric neutrality and beta equilibrium do not strongly affect the size of the inhomogeneous window,
other effects, like vector interactions or the presence of background magnetic fields, enlarge it significantly.
The role of color superconductivity, on the other hand, is not yet completely clear. 
According to NJL-model studies in isospin symmetric matter~\cite{Sadzikowski:2002iy},
inhomogeneous chiral condensates may coexist with a 
homogeneous diquark condensate if the diquark coupling is small, but become suppressed for larger couplings.
This could change, however, if electric neutrality is imposed,
since it  strongly disfavors the BCS pairing of up and down quarks. 
As a possible result, the diquark condensates could become spatially modulated as well or be suppressed completely.  
In any case, this
implies that, if quark matter is present inside the core of these stars, it is most likely in a crystalline state.
In particular, the realization of a coexistence phase where both color-superconducting and chiral condensates 
are inhomogeneous is a fascinating possibility, which might be worth investigating.

Hybrid stars with an inhomogeneous quark matter-core are compatible (within model uncertainties) with the recent measurements of pulsars with 2 solar masses  \cite{Carignano:2015kda}. Since the presence of crystalline phases 
has however only a small influence on the equation of state,
their existence or non-existence cannot be concluded on the basis of mass measurements, 
even in conjunction with precisely determined compact-star radii. 
Indeed, model uncertainties and systematic errors such as those introduced by employing spherically-symmetric TOV equations 
in an anisotropic system are expected to dominate over these effects within this type of calculation.
It is therefore highly desirable to work out more sensitive signatures of their formation.
A first example in this direction is the neutrino emissivity, discussed in \cite{Tatsumi:2014cea},  
which is related to the cooling properties of the star.
A complete calculation of the cooling process of the whole star would however be required before a detailed comparison with astrophysical data could be made. 

The formation of inhomogeneous phases might also have a stronger influence on the transport properties of matter inside compact stars, particularly through their peculiar low-energy excitations, such as phonons. Some work on calculating the dispersion relations for these excitations has been performed in \cite{Lee:2015bva} for the case of a CDW modulation and \cite{Hidaka:2015xza} for a RKC. 
It would also be interesting to investigate phenomena like glitches and the emission of gravitational waves,
which have been investigated for crystalline color superconductors (cf.~ref.~\cite{Anglani:2013gfu}),
but not yet for the specific case of inhomogeneous chiral condensates.

\section*{Acknowledgment}
S.C. acknowledges financial support by the BMBF under grant no.~05P12RDGHD.


\begin{thebibliography}{68}

\bibitem{Ivanenko:1965dg}
D.D. Ivanenko, D.F. Kurdgelaidze, Astrophysics \textbf{1}, 251 (1965),
  [Astrofiz.1,479(1965)]

\bibitem{Itoh:1970uw}
N.~Itoh, Prog. Theor. Phys. \textbf{44}, 291 (1970)

\bibitem{Demorest}
P.~Demorest, T.~Pennucci, S.~Ransom, M.~Roberts, J.~Hessels, Nature
  \textbf{467}, 1081 (2010), \texttt{arXiv:1010.5788}

\bibitem{Antoniadis}
J.~Antoniadis et~al., Science \textbf{340}, 6131 (2013),
  \texttt{arXiv:1304.6875}

\bibitem{Alford:2006vz} 
M.~Alford, D.~Blaschke, A.~Drago, T.~Kl{\"a}hn, G.~Pagliara, J.~Schaffner-Bielich,
  Nature \textbf{445}, E7 (2007),
  \texttt{arXiv:astro-ph/0606524}

\bibitem{Alford:2015gna} 
  M.~G.~Alford, S.~Han,
  \texttt{arXiv:1508.01261}

\bibitem{Buballa:2014jta} 
  M.~Buballa {\it et al.},
  J.\ Phys.\ G \textbf{41},123001 (2014),
  \texttt{arXiv:1402.6911}

\bibitem{Kojo:2009}
T.~Kojo, Y.~Hidaka, L.~McLerran, R.~Pisarski, Nucl. Phys. A \textbf{843}, 37
  (2010), \texttt{arXiv:0912.3800}

\bibitem{Muller:2013tya}
D.~M{\"u}ller, M.~Buballa, J.~Wambach, Phys. Lett. B \textbf{727}, 240 (2013),
  \texttt{arXiv:1308.4303}

\bibitem{CB:2011}
S.~Carignano, M.~Buballa, Acta Phys. Polon. Supp. \textbf{5}, 641 (2012),
  \texttt{arXiv:1111.4400}

\bibitem{Anglani:2013gfu}
R.~Anglani, R.~Casalbuoni, M.~Ciminale, N.~Ippolito, R.~Gatto et~al., Rev. Mod.
  Phys. \textbf{86}, 509 (2014), \texttt{arXiv:1302.4264}

\bibitem{Buballa:2014tba}
M.~Buballa, S.~Carignano, Prog. Part. Nucl. Phys. \textbf{81}, 39 (2015),
  \texttt{arXiv:1406.1367}

\bibitem{Broniowski:2011}
W.~Broniowski, Acta Phys. Polon. Supp. \textbf{5}, 631 (2012),
  \texttt{arXiv:1110.4063}

\bibitem{Nickel:2009wj}
D.~Nickel, Phys. Rev. D \textbf{80}, 074025 (2009), \texttt{arXiv:0906.5295}

\bibitem{NJL2}
Y.~Nambu, G.~Jona-Lasinio, Phys. Rev. \textbf{124}, 246 (1961)

\bibitem{Abuki:2011}
H.~Abuki, D.~Ishibashi, K.~Suzuki, Phys.Rev. D \textbf{85}, 074002 (2012),
  \texttt{arXiv:1109.1615}

\bibitem{Carignano:2012sx}
S.~Carignano, M.~Buballa, Phys. Rev. D \textbf{86}, 074018 (2012),
  \texttt{arXiv:1203.5343}

\bibitem{Broniowski:1990}
W.~Broniowski, A.~Kotlorz, M.~Kutschera, Acta Phys. Polon. B \textbf{22}, 145
  (1991)

\bibitem{NT:2004}
E.~Nakano, T.~Tatsumi, Phys. Rev. D \textbf{71}, 114006 (2005)

\bibitem{Schnetz:2004}
O.~Schnetz, M.~Thies, K.~Urlichs, Annals Phys. \textbf{314}, 425 (2004)

\bibitem{CNB:2010}
S.~Carignano, D.~Nickel, M.~Buballa, Phys. Rev. D \textbf{82}, 054009 (2010),
  \texttt{arXiv:1007.1397}

\bibitem{Buballa:2012vm}
M.~Buballa, S.~Carignano, Phys. Rev. D \textbf{87}, 054004 (2013),
  \texttt{arXiv:1210.7155}

\bibitem{Frolov:2010}
I.~Frolov, V.~Zhukovsky, K.~Klimenko, Phys.Rev. D \textbf{82}, 076002 (2010),
  \texttt{arXiv:1007.2984}

\bibitem{Nickel:2009ke}
D.~Nickel, Phys. Rev. Lett. \textbf{103}, 072301 (2009),
  \texttt{arXiv:0902.1778}

\bibitem{Abuki:2013vwa}
H.~Abuki, Phys. Rev. D \textbf{87}, 094006 (2013), \texttt{arXiv:1304.1904}

\bibitem{Abuki:2013pla}
H.~Abuki, Phys. Lett. B \textbf{728}, 427 (2014), \texttt{arXiv:1307.8173}

\bibitem{Nowakowski:2015ksa}
D.~Nowakowski, M.~Buballa, S.~Carignano, J.~Wambach, Proceedings of CSQCD IV
  (2015), \texttt{arXiv:1506.04260}

\bibitem{Asakawa:1989}
M.~Asakawa, K.~Yazaki, Nucl. Phys. A \textbf{504}, 668 (1989)

\bibitem{Frank:2003ve}
M.~Frank, M.~Buballa, M.~Oertel, Phys. Lett. B \textbf{562}, 221 (2003),
  \texttt{arXiv:hep-ph/0303109}

\bibitem{dno}
D.~Nowakowski, M.~Buballa, S.~Carignano, In preparation  (2015)

\bibitem{Lai:1991}
D.~Lai, S.~Shapiro, Astrophys. J. \textbf{383}, 745 (1991)

\bibitem{Ferrer:2010wz}
E.J. Ferrer, V.~de~la Incera, J.P. Keith, I.~Portillo, P.L. Springsteen, Phys.
  Rev. C \textbf{82}, 065802 (2010), \texttt{arXiv:1009.3521}

\bibitem{Bocquet} 
M.~Bocquet, S.~Bonazzola, E.~Gourgoulhon, J.~Novak,
Astron.\ Astrophys.\  \textbf{301}, 757 (1995), \texttt{arXiv:gr-qc/9503044}
 
\bibitem{Cardall} 
C.~Y.~Cardall, M.~Prakash, J.~M.~Lattimer,
Astrophys.\ J.\  \textbf{554}, 322 (2001),
\texttt{astro-ph/0011148}

\bibitem{suga1}
H.~Suganuma, T.~Tatsumi, 
Annals Phys. \textbf{208}, 470 (1989)

\bibitem{Gusyin:1994}
V.P. Gusynin, V.A. Miransky, I.A. Shovkovy, Phys. Rev. Lett. \textbf{73}, 3499
  (1994)

\bibitem{Gusynin:1995nb}
V.P. Gusynin, V.A. Miransky, I.A. Shovkovy, Nucl. Phys. B \textbf{462}, 249
  (1996), \texttt{arXiv:hep-ph/9509320}

\bibitem{Miransky:2015ava}
V.A. Miransky, I.A. Shovkovy, Phys. Rept. \textbf{576}, 1 (2015),
  \texttt{arXiv:1503.00732}

\bibitem{Andersen:2014xxa}
J.O. Andersen, W.R. Naylor, A.~Tranberg (2014), \texttt{arXiv:1411.7176}

\bibitem{Schon:2000he}
V.~Sch{\"o}n, M.~Thies, Phys. Rev. D \textbf{62}, 096002 (2000),
  \texttt{arXiv:hep-th/0003195}

\bibitem{Basar:2009fg}
G.~Basar, G.V. Dunne, M.~Thies, Phys. Rev. D \textbf{79}, 105012 (2009)

\bibitem{tatplb15}
 T.~Tatsumi, K.~Nishiyama, S.~Karasawa,
 Phys. Lett. B \textbf{743}, 66 (2015) 

\bibitem{Nishiyama:2015fba}
K.~Nishiyama, S.~Karasawa, T.~Tatsumi, 
Phys. Rev. D \textbf{92}, 036008 (2015), \texttt{arXiv:1505.01928}

\bibitem{Walecka:1974qa}
J.D. Walecka, Annals Phys. \textbf{83}, 491 (1974)

\bibitem{Vogl:1991qt}
U.~Vogl, W.~Weise, Prog. Part. Nucl. Phys. \textbf{27}, 195 (1991)

\bibitem{Ebert:1985kz}
D.~Ebert, H.~Reinhardt, Nucl. Phys. B \textbf{271}, 188 (1986)

\bibitem{Steinheimer:2010sp}
J.~Steinheimer, S.~Schramm, Phys. Lett. B \textbf{696}, 257 (2011),
  \texttt{arXiv:1005.1176}

\bibitem{Steinheimer:2014kka}
J.~Steinheimer, S.~Schramm (2014), \texttt{arXiv:1401.4051}

\bibitem{Marco:MSc}
M.~Schramm, Master's thesis, TU Darmstadt (2013)

\bibitem{Zhang:2009mk}
Z.~Zhang, T.~Kunihiro, Phys. Rev. D \textbf{80}, 014015 (2009)

\bibitem{Abuki:2009ba}
H.~Abuki, R.~Gatto, M.~Ruggieri, Phys. Rev. D \textbf{80}, 074019 (2009),
  \texttt{arXiv:0904.0866}

\bibitem{Shuster}
E.~Shuster, D.T. Son, Nucl. Phys. B \textbf{573}, 434 (2000)

\bibitem{Berges:1998rc}
J.~Berges, K.~Rajagopal, Nucl. Phys. B \textbf{538}, 215 (1999),
  \texttt{arXiv:hep-ph/9804233}

\bibitem{Schwarz:1999dj}
T.~Schwarz, S.~Klevansky, G.~Papp, Phys. Rev. C \textbf{60}, 055205 (1999),
  \texttt{arXiv:nucl-th/9903048}

\bibitem{Buballa:2005}
M.~Buballa, Phys. Rept. \textbf{407}, 205 (2005), \texttt{arXiv:hep-ph/0402234}

\bibitem{Muller:2013pya}
D.~M{\"u}ller, M.~Buballa, J.~Wambach, Eur. Phys. J. A \textbf{49}, 96 (2013),
  \texttt{arXiv:1303.2693}

\bibitem{Sadzikowski:2002iy}
M.~Sadzikowski, Phys. Lett. B \textbf{553}, 45 (2003),
  \texttt{arXiv:hep-ph/0210065}

\bibitem{Sadzikowski:2006jq}
M.~Sadzikowski, Phys. Lett. B \textbf{642}, 238 (2006),
  \texttt{arXiv:hep-ph/0609186}

\bibitem{Klahn:2006iw}
T.~Kl{\"a}hn, D.~Blaschke, F.~Sandin, C.~Fuchs, A.~Faessler, H.~Grigorian,
  G.~R{\"o}pke, J.~Tr{\"u}mper, Phys. Lett. B \textbf{654}, 170 (2007),
  \texttt{arXiv:nucl-th/0609067}

\bibitem{Alford:2002kj}
M.~Alford, K.~Rajagopal, JHEP \textbf{0206}, 031 (2002),
  \texttt{arXiv:hep-ph/0204001}

\bibitem{Giannakis:2004pf}
I.~Giannakis, H.C. Ren, Phys. Lett. B \textbf{611}, 137 (2005),
  \texttt{arXiv:hep-ph/0412015}

\bibitem{Rajagopal:2000ff}
K.~Rajagopal, F.~Wilczek,
Phys. Rev. Lett. \textbf{86}, 3492 (2001),
\texttt{arXiv:hep-ph/0012039}

\bibitem{Ruester:2005jc} 
S.B.~Ruester, V.~Werth, M.~Buballa, I.A.~Shovkovy, D.H.~Rischke,
Phys. Rev. D \textbf{72}, 034004 (2005),
\texttt{arXiv:hep-ph/0503184}

\bibitem{Abuki:2004zk} 
H.~Abuki, M.~Kitazawa, T.~Kunihiro,
Phys. Lett. B \textbf{615}, 102 (2005),
\texttt{arXiv:hep-ph/0412382}

\bibitem{Baldo:2002ju} 
M.~Baldo, M.~Buballa, F.~Burgio, F.~Neumann, M.~Oertel, H.~J.~Schulze,
Phys. Lett. B {\bf 562}, 153 (2003),
\texttt{arXiv:nucl-th/0212096}

\bibitem{Buballa:2003et} 
M.~Buballa, F.~Neumann, M.~Oertel, I.~Shovkovy,
Phys. Lett. B {\bf 595}, 36 (2004),
\texttt{arXiv:nucl-th/0312078}

\bibitem{Moreira:2013ura}
J.~Moreira, B.~Hiller, W.~Broniowski, A.~Osipov, A.~Blin, Phys. Rev. D
  \textbf{89}, 036009 (2014), \texttt{arXiv:1312.4942}

\bibitem{Carignano:2015kda}
S.~Carignano, E.J.~Ferrer, V.~de~la Incera, L.~Paulucci, Phys. Rev. D 
 \textbf{92}, 105018 (2015),  \texttt{arXiv:1505.05094}

\bibitem{Tatsumi:2014cea}
T.~Tatsumi, T.~Muto, Phys. Rev. D \textbf{89}(10), 103005 (2014),
  \texttt{arXiv:1403.1927}

\bibitem{Oppenheimer:1939ne}
J.R.~Oppenheimer, G.M.~Volkoff, Phys. Rev. \textbf{55}, 374 (1939)

\bibitem{Tolman:1939jz}
R.C.~Tolman, Phys. Rev. \textbf{55}, 364 (1939)

\bibitem{Chatterjee:2014}
D.~Chatterjee, T.~Elghozi, J.~Novak, M.~Oertel,
Mon. Not. Roy. Astron. Soc. \textbf{447}, 3785  (2015),
 \texttt{arXiv:1410.6332}

\bibitem{Heinz:2013hza}
A.~Heinz, F.~Giacosa, D.H.~Rischke, Nucl. Phys. A \textbf{933}, 34 (2015),
  \texttt{arXiv:1312.3244}

\bibitem{Pagliara:2007ph}
G.~Pagliara, J.~Schaffner-Bielich, Phys. Rev. D \textbf{77}, 063004 (2008),
  \texttt{arXiv:0711.1119}

\bibitem{Glendenning:1992vb} 
N.~K.~Glendenning, Phys. Rev. D {\bf 46}, 1274 (1992)

\bibitem{Alkofer:1994ph}
R.~Alkofer, H.~Reinhardt, H.~Weigel, Phys. Rept. {\bf 265}, 139 (1996),
\texttt{[arXiv:hep-ph/9501213]}

\bibitem{Christov:1995vm}
C.~Christov et~al., Prog. Part. Nucl. Phys. {\bf 37} 91 (1996),
\texttt{[arXiv:hep-ph/9604441]}

\bibitem{Ripka}
G.~Ripka,
Quarks Bound by Chiral Fields: The Quark Structure of the
Vacuum and of Light Mesons and Baryons, 
Oxford Studies in Nuclear Physics. Clarendon Press (1997)

\bibitem{Pavlov:Vela}
G.G.~Pavlov, V.E.~Zavlin, D.~Sanwal, V.~Burwitz, G.P.~Garmire,
  Astrophys. J. \textbf{552}, L129 (2001)

\bibitem{Slane:3C58}
P.O.~Slane, D.J.~Helfand, S.S.~Murray, Astrophys. J.
  \textbf{571}, L45 (2002)

\bibitem{Ho:CasA}
W.C.G.~Ho, C.O.~Heinke, Nature \textbf{462}(7269), 71 (2009),
  \texttt{http://dx.doi.org/10.1038/nature08525}

\bibitem{Iwamoto:1980eb}
N.~Iwamoto, Phys. Rev. Lett. \textbf{44}, 1637 (1980)

\bibitem{Maxwell:1977zz}
O.~Maxwell, G.E.~Brown, D.K.~Campbell, R.F.~Dashen, J.T.~Manassah, Astrophys.
  J. \textbf{216}, 77 (1977)

\bibitem{Lee:2015bva}
T.G.~Lee, E.~Nakano, Y.~Tsue, T.~Tatsumi, B.~Friman (2015),
  \texttt{arXiv:1504.03185}

\bibitem{Hidaka:2015xza}
Y.~Hidaka, K.~Kamikado, T.~Kanazawa, T.~Noumi, Phys. Rev. D \textbf{92}(3),
  034003 (2015), \texttt{arXiv:1505.00848}

\end{thebibliography}
\end{document}